# Emergent Phenomena with Broken Parity-Time Symmetry: Odd-order vs. Even-order Effects

Sang-Wook Cheong* and Fei-Ting Huang
Rutgers Center for Emergent Materials and Department of Physics and Astronomy, Rutgers University
*Corresponding author: sangc@physics.rutgers.edu

**ABSTRACT**
**Symmetry often governs the laws of nature, and breaking symmetry accompanies a new order parameter and emergent observable phenomena. Herein, we focus on broken Parity (P)-Time (T) symmetry, which lifts the Kramers' degeneracy, and thus, guarantees non-trivial spin textures in excitation spectra. To attain non-zero measurables, we use the concept of symmetry operational similarity (SOS), which consider the symmetry relationship between a specimen and an experimental setup, rather than the symmetry of specific coupling terms. Even without specific coupling terms, this SOS approach can tell if the relevant phenomenon is a zero, non-zero odd-order or non-zero even-order effect. We discuss systematically numerous steady-state physical phenomena, in which breaking P-T symmetry is a necessary condition. These phenomena include Odd-order or Even-order Anomalous Hall Effect, Optical activities, Directional nonreciprocity in transverse magnetic field, Diagonal or Off-diagonal current-induced magnetization (current can be associated with electrons, phonons, or light), Diagonal or Off-diagonal piezomagnetism and piezoelectricity. Some of these phenomena turn out to be conjugate to each other through P ↔T. Our findings unveil numerous new non-traditional candidate materials for various exotic physical phenomena, many of which have never been realized in the standard coupling term/tensorial approaches, and are a transformative and unconventional avenue for symmetry-guided materials designs and discoveries.**

## I. INTRODUCTION

Symmetries have long played a crucial role in nature. Two fundamental symmetries are, space inversion (**I** or Parity **P**), changing the sign of the spatial coordinates of the four vector $(x,y,z,t) \rightarrow (-x,-y,-z,t)$ as our right-left hand symmetry and time reversal symmetry (**T**), changing the sign of the time component of the four vector: $(x,y,z,t) \rightarrow (x,y,z,-t)$ [1]. For many years, it was thought that nature exhibits **P** and **T** symmetries. The discovery of the breakdown of the parity conservation of nature led to Lee and Yang being awarded the Nobel prize in 1957 [2]. 1980 Nobel prize was awarded to Cronin and Fitch for the finding that nature excludes the time reversal invariance[3]. Later on, the papers by Bender and Boettcher in 1998 [4, 5] play a crucial role in bringing **P-T** symmetry to the forefront of quantum mechanics research, leading to further investigations and developments in the field [5-8]. Herein, **I**⊗**T**, (the combination of **I** and **T**, symbolized as **I**⊗**T**), is used for the **P-T** symmetry operation. The prominent implication of **I**⊗**T** symmetry breaking is to create nontrivial physics by lifting the Kramers degeneracy and non-zero magnetic textures of, for example, electronic band structures.

By considering an excitation with wave vector (velocity or linear momentum), which is magnetically polarized. This excitation is associated with ***k*** and a combination of magnetization ***M*** with any arbitrary relative angle between ***k*** and ***M***. Then, the **I**⊗**T** symmetry operation links ($k$,$M$) and ($k$,-$M$). In the presence of $I \otimes T$ symmetry, all bulk bands are doubly degenerate at all momenta in the Brillouin zone. Thus, the magnetic polarization of the excitation requires broken $I \otimes T$ symmetry. In other words, non-zero magnetic textures of electronic band structures or phonon spectra require broken $I \otimes T$ symmetry [9-11]. For example, the spin degeneracy of surface bands can be lifted by the spin-orbit interaction because of space-inversion symmetry is broken at the terminated surface [12]. The spin-momentum locking of the topological surface states of topological insulators and Weyl semimetal, also that of the surface or bulk Rashba states all originate from broken $I \otimes T$ symmetry [9-11]. Magnetically polarized phonons are called phonons with kinetic magnetoelectricity, polarized phonons, or chiral phonons [13-15]. The term chiral phonon is often used, but the magnetically polarized phonons are not chiral (i.e., some mirror symmetries with spatial rotations are unbroken), so the term chiral phonons is not appropriate. Emphasize that the requirement for magnetically polarized phonons is broken $I \otimes T$, which has no direction.

Here, we explore the magnetic point group-property connection in the broken $I \otimes T$ class, utilizing the symmetry operational similarity (SOS) approach in the same spirit as the known Curie Principle[16] and Neumann's Principle[17], while the group symmetry of the whole experimental setup is now considered. When a specimen has SOS with "the whole experimental setup with specific specimen environments and measuring probes to measure a particular phenomenon/property", the relevant physical effect can be non-zero in the specimen [18-21]. More precisely speaking, SOS means that the specimen has equal or lower, but not higher symmetries, compared with the whole experimental setup. This SOS approach is more general than the traditional approach of considering the symmetry of specific coupling (tensorial) terms and can tell if the relevant phenomenon is a zero, non-zero odd-order or non-zero even-order effect, since the sign of the relevant parameter (for example, the measurable) can be readily considered. In the SOS consideration, the response of the specimen-to-specimen environments does not have to be considered a priori but is directly relevant to the measuring phenomenon/property and can be obtained from the SOS relationship. We conclude that specimens with broken $I \otimes T$ symmetry can possess some or all the following properties: Odd-order or Even-order anomalous Hall effect (AHE), Diagonal/Off-diagonal piezomagnetism, Directional nonreciprocity in transverse magnetic field ($H$), Diagonal/Off-diagonal Odd-order or Even-order current-induced magnetization, Optical activities, and Diagonal/Off-diagonal piezoelectricity. We have made clear connections among these seemingly distinct phenomena through broken $I \otimes T$ symmetry. Note that all these phenomena with broken $I \otimes T$ symmetry are associated with the spin textures in excitation spectra, including the ground state, but their specific relationship is beyond the scope of this work. Figure 1 summarizes the inter-relationships between physical properties and magnetic point groups (MPGs), which is a graphical representation of the SOS relationships [19, 21]. 122 magnetic point groups have been well-defined and readers are encouraged to the provided references [22, 23] [24] for a more comprehensive understanding. Taking the middle of the diagram, there are 16 MPGs allowing all the above phenomena. Three main divisions that separates broken $I \otimes T$ symmetry MPGs are the ferromagnetic, polar, and chiral groups. It is partially overlapped with the magic trinity published by Schmid[25] that contains ferromagnetism, ferroelectricity and ferrotoroidic fields. Those exceptions in each group are emphasized and will be discussed later.

The approach with coupling terms in Hamiltonian can be associated with various ranks of tensors[26, 27]. Space-time reflection symmetry restrictions of the symmetry-adapted tensors have been examined and classified by considering magnetic symmetry groups through tensor approaches[27-35] or multipole classification [36]. The results of our SOS approach overlap with those obtained from the standard tensorial approach. Moreover, we demonstrate several "forbidden" cases based on standard tensor approaches while they can exhibit high-order effects due to the symmetry broken by experimental setups. We also identify those cases when behaviors along *x*/*y* are different those along *xy*/*yx*. Our SOS approach serves an alternative physically intuitive way to design experiments prior to the requirement of constitutive coupling terms and different ranks of tensors. The microscopic mechanism for physical phenomena and their magnitudes are beyond the scope of our symmetry discussion.

## II. METHODS

We have analyzed all broken symmetries of "the whole experimental setup with specific specimen environments and measuring probes to measure a particular phenomenon". The considered odd-order or even-order phenomena include AHE, optical activities, diagonal current-induced magnetization, off-diagonal current-induced magnetization, diagonal piezomagnetism, off-diagonal piezomagnetism, and directional nonreciprocity in transverse magnetic field. We have considered the symmetry operations of spatial rotations, mirror reflections, space inversion, and time reversal. We have allowed free translation for all cases, and free spatial rotations along the one-dimensional (1D) direction for 1D phenomena such as optical activities, diagonal piezomagnetism, and diagonal current-induced magnetization. After the analysis for each experimental setup, we have precisely identified all MPGs that have SOS with the experimental setup with a certain set of broken symmetries. SOS means that the magnetic groups have equal or lower, but not higher symmetries, compared with the experimental setup.

We have considered symmetries that are along only those basis vectors of the conventional crystallographic coordinate systems. Those basis vectors are given in the settings of 122 MPGs [24, 37]. For example, we discuss the symmetries along *x*, *y*, *z*, *xy*, and *yx* directions of the tetragonal and cubic MPGs while only symmetries along *x*, *y*, *z* directions in orthorhombic MPGs are considered in SOS analysis. Even for hexagonal and trigonal structures, *x*, *y*, and *z* are defined to be orthogonal to each other. The rotation $C_3$ encompasses both $C_3^+$ (counterclockwise) and $C_3^-$ (clockwise), with the + and – signs. Any symmetry criteria associated with $C_3$ are equally applicable to the operation on $C_3^+$ and $C_3^-$, concurrently. This principle extends to $C_4$ and $C_6$ rotations. The analysis of 122 MPGs is summarized in Fig. 1. While maintaining the focus on these 122 groups for simplicity, we have expanded Table 1 to include 190 magnetic point groups to better assist readers in precisely locating physical properties. This update thoughtfully addresses directional considerations in various settings, providing additional clarity. For instance, in the monoclinic system, we now distinguish between $2'_{[x]}$, $2'_{[y]}$, and $2'_{[z]}$, denoting the unique axis along the *x*- or *a*-axis, *y*- or *b*-axis and *z*- or *c*-axis, respectively. Similarly, for orthorhombic system, e.g. *mm*2, 2*mm*, and *m*2*m*, different settings are considered. In the trigonal system, we have listed MPG 3*m* separately by 3*m*1 with $M_x=m_{100}$ and 31*m* with $M_y=m_{210}$.

## III. RESULTS AND DISCUSSION
### A. Symmetry of the Odd-order AHE measurements

The Hall effect, a hallmark of how Maxwell's equations work in materials, was discovered by Edwin Hall while he was working on his doctoral degree in 1879, and has been well utilized to measure carrier density as well as detect small magnetic fields [38]. This so-called ordinary Hall effect contrasts with the anomalous Hall effect (AHE) in "ferromagnets", which is sometimes called extraordinary Hall effect [39]. This AHE exists in zero applied magnetic field, and varies linearly with applied electric current, so its sign changes when the current direction is reversed. It was proposed that AHE can exist in "truly antiferromagnetic" systems such as $Mn_3$(Rh,Ir,Pt) with Kagome lattice [40, 41], originating from the Berry curvature. In fact, $Mn_3$(Sn,Ge), forming in the same crystallographic structure with that of $Mn_3$(Rh,Ir,Pt), is experimentally reported to exhibit a significant AHE [42, 43]. However, it turns out that $Mn_3$(Sn,Ge) does exhibit a small, but finite net magnetic moment [44]. The precise experimental characterization of $Mn_3$(Rh,Ir,Pt) was unclear until a recent report [45], partially due to the presence of competing multiple magnetic states in the system. Topological Hall effects in skyrmion systems have been reported [46, 47], and occur typically in the presence of external magnetic fields. Nonlinear (quadratic-type) AHE has been also observed [48-51] in, e.g., bilayer $T_d$-$WTe_2$. [49] This quadratic AHE varies with the square of current, so its sign does not change when the current direction is reversed. Note that AHE of antiferromagnets without any net moment can be particularly useful for the fast sensing of magnetic fields due to the intrinsic fast dynamics of antiferromagnets [52].

Herein, we define AHE as "Transverse voltage induced by applied current in zero magnetic field". The sets of (electric current, +/−), (electric current, h/c), (thermal current, +/ −), and (thermal current, h/c) correspond to the Hall, Ettingshausen, Nernst, and thermal Hall effects, all of which we call Hall-type effects, respectively (+/- means an induced voltage difference and h/c (hot/cold) means an induced thermal gradient, accumulated on off-diagonal surfaces). In terms of symmetry, there is no difference of any (applied) constant quasi-participle current (electric current, phonon current or spin wave current). In addition, there is no difference between (induced) temperature gradient and (induced) voltage gradient in terms of symmetry. Thus, there is little difference among the requirements for these four types of Hall effects as long as the relevant electric/thermal current can exist, so, for example, the existence of non-zero AHE means the presence of non-zero anomalous Nernst effect. With this multi-faceted nature of Hall-type effects, it is imperative to find the accurate relationship among all different kinds of Hall-type effects, and also the requirements to have non-zero values of various Hall-type effects.

To find the requirements of broken symmetries for various phenomena, we, first, define the general symmetry operation notations for three orthogonal $x$, $y$, and $z$ axes such as $\mathbf{R}_x$=2-fold rotation along the $x$ axis, $\mathbf{M}_x$=mirror reflection with mirror perpendicular to the $x$ axis, $\mathbf{I}$=space inversion, $\mathbf{T}$=time reversal, etc. Then, we have these general relationships: $\mathbf{R}_x \otimes \mathbf{R}_y = \mathbf{R}_z$, $\mathbf{M}_z \otimes \mathbf{R}_y = \mathbf{M}_x$, $\mathbf{M}_x \otimes \mathbf{R}_z = \mathbf{M}_y$, $\mathbf{M}_x \otimes \mathbf{R}_y = \mathbf{M}_z$, $\mathbf{M}_y \otimes \mathbf{M}_z = \mathbf{R}_x$, $\mathbf{M}_x \otimes \mathbf{M}_z = \mathbf{R}_y$, $\mathbf{M}_x \otimes \mathbf{M}_y = \mathbf{R}_z$, $\mathbf{M}_x \otimes \mathbf{R}_x = \mathbf{M}_y \otimes \mathbf{R}_y = \mathbf{M}_z \otimes \mathbf{R}_z = \mathbf{I}$, $\mathbf{M}_x = \mathbf{R}_x \otimes \mathbf{I}$, $\mathbf{M}_y = \mathbf{I} \otimes \mathbf{R}_y$, and $\mathbf{M}_z = \mathbf{I} \otimes \mathbf{R}_z$ (all of these are commutative.). All of our whole experimental setups have translational symmetry, so we can disregard or freely allow any translations. Similarly, when we consider experimental setups invariant under any rotations along a particular direction, then we disregard or freely allow any rotations along the axis.

We can have these transformations for the experimental setup for AHE measurements in Figs. 2(a)-(d): (a) ↔ (a) by $\mathbf{M}_z$, $\mathbf{M}_x \otimes \mathbf{T}$, and $\mathbf{R}_y \otimes \mathbf{T}$; (a) ↔ (d) by $\mathbf{I}$, $\mathbf{R}_z$, $\mathbf{M}_y \otimes \mathbf{T}$, and $\mathbf{R}_x \otimes \mathbf{T}$; (a) ↔ (b) by $\mathbf{I} \otimes \mathbf{T}$, $\mathbf{M}_y$, $\mathbf{R}_x$, and $\mathbf{R}_z \otimes \mathbf{T}$; (a) ↔ (c) by $\mathbf{T}$, $\mathbf{M}_x$, $\mathbf{R}_y$, and $\mathbf{M}_z \otimes \mathbf{T}$; Thus, Odd-order $AHE_{yx}$ means Odd-order AHE with current along $x$ and Hall voltage along $y$ in Fig. 2(a), and the experimental setup to measure odd-order $AHE_{yx}$ has unbroken [$\mathbf{1},\mathbf{I},\mathbf{M}_z,\mathbf{R}_z,\mathbf{M}_x \otimes \mathbf{T},\mathbf{M}_y \otimes \mathbf{T},\mathbf{R}_x \otimes \mathbf{T},\mathbf{R}_y \otimes \mathbf{T}$] from Fig. 2(a) ↔ (a) and (a) ↔ (d) and broken {$\mathbf{I} \otimes \mathbf{T},\mathbf{T},\mathbf{M}_x,\mathbf{M}_y,\mathbf{R}_x,\mathbf{R}_y,\mathbf{C}_{3x},\mathbf{M}_z \otimes \mathbf{T},\mathbf{R}_z \otimes \mathbf{T}$} from Fig. Fig. 2(a) ↔ (b) and (a) ↔ (c). Note that unbroken [$\mathbf{1},\mathbf{I},\mathbf{M}_z,\mathbf{R}_z,\mathbf{M}_x \otimes \mathbf{T},\mathbf{M}_y \otimes \mathbf{T},\mathbf{R}_x \otimes \mathbf{T},\mathbf{R}_y \otimes \mathbf{T}$] corresponds to MPG of $m'm'm$, and Odd-order is in terms of current ($J$), i.e., goes like $V_1 J + V_3 J^3 + V_5 J^5 + - --$ ($V_i$'s are constants). Now, any specimens having SOS with the above experimental setup will show Odd-order AHE$_{yx}$ when they have broken {**I⊗T,T,M$_x$,M$_y$,R$_x$,R$_y$,C$_{3x}$,M$_z$⊗T,R$_z$⊗T**}([1] in TABLE 1). The rest independent ones can be either broken or unbroken. For example, for broken {**I**}, basically, Odd-order AHE of the original domain is same with that of the domain after space inversion. Emphasize that the requirements for anomalous Ettingshausen, anomalous Nernst, and anomalous thermal Hall effects are identical for those for AHE. It turns out that all ferromagnetic point groups [25, 34, 53] can have non-zero net magnetic moments, and do have broken {**I⊗T,T,M$_x$,M$_y$,R$_x$,R$_y$,C$_{3x}$,M$_z$⊗T,R$_z$⊗T**} when the net magnetic moments are along $z$. In general, the presence of Odd-order AHE$_{yx}$ does not necessarily mean non-zero Odd-order AHE$_{xy}$; AHE$_{xy}$. For example, if there is broken C$_{3y}$ and unbroken C$_{3x}$, then AHE$_{xy}$ can be non-zero even though AHE$_{yx}$ should be zero. However, in all ferromagnetic point groups that can have non-zero net moments along $z$, Odd-order AHE$_{yx}$ is same with Odd-order AHE$_{xy}$ except their sign difference (i.e., they are anti-symmetric).

In all ferromagnetic point groups that can have non-zero net moments along $z$ ($x$ or $y$), Odd-order AHE$_{yx}$ (AHE$_{zy}$ or AHE$_{zx}$) is same with Odd-order AHE$_{xy}$ (AHE$_{yz}$ or AHE$_{xz}$) except their sign difference (i.e. they are anti-symmetric). However, there are three types of non-ferromagnetic point groups which allow Odd-order AHE; (1) MPGs of C$_{3z}$ can have Odd-order AHE$_{zx}$ or AHE$_{zy}$ with broken {**I⊗T,T,M$_z$,M$_y$,R$_z$,R$_y$,C$_{3y}$,M$_x$⊗T,R$_x$⊗T**}, but they have zero Odd-order AHE$_{xz}$ or AHE$_{yz}$ due to unbroken C$_{3z}$: the examples are 3, $\bar{3}$, 32, 3$m$, $\bar{3}m$, 6', $\bar{6}$', 6'/$m$', 6'22', 6'$mm$', $\bar{6}$'$m$2', $\bar{6}$'2$m$', 6'/$m$'$mm$' for Odd-order AHE$_{zy}$ and 3, $\bar{3}$, 32', 3$m$', $\bar{3}m$', 6', $\bar{6}$', 6'/$m$', 6'$m$'$m$, 6'2'2, $\bar{6}$'2'$m$, $\bar{6}$'$m$'2, 6'/$m$'$m$'$m$ for Odd-order AHE$_{zx}$. (2) MPGs of C$_{4z}$⊗T or C$_{4z}$⊗I⊗T can have Odd-order AHE$_{yx}$ and Odd-order AHE$_{xy}$ with broken{**I⊗T,T,M$_x$,M$_y$,R$_x$,R$_y$,C$_{3x}$,C$_{3y}$,M$_z$⊗T,R$_z$⊗T**}: the examples are 4', 4', $\bar{4}$', 4'/$m$, 4'2'2, 4'$m$'$m$, $\bar{4}$'$m$2, $\bar{4}$'2'$m$, and 4'/$mm$'$m$. (3) Cubic MPGs allow Odd-order AHE$_{yx,xy}$ as well as Odd-order AHE$_{xy,yx}$: 23, $m\bar{3}$, 4'32', $\bar{4}$'3$m$', and $m\bar{3}m$'. The finalized lists are shown in [1], [3], [5], [7], and [9] within TABLE 1.

All inversion-symmetric spin order on Kagome lattice in Figs. 3(a)-(d) do have broken {**I⊗T,T,M$_x$,M$_y$,R$_x$,R$_y$,C$_{3x}$,M$_z$⊗T,R$_z$⊗T**}, and Mn$_3$(Sn,Ge,Ga,Rh,Ir,Pt) with these spin orders, in fact, all belong to ferromagnetic point groups, and are supposed to exhibit odd-order AHE$_{yx}$ and AHE$_{xy}$ (the relevant $z$-axis is along the ***M*** direction shown in Figs. 3(a)-(d)). Note that all Figs. 3(a)-(d) have SOS with ***M***, so they can exhibit non-zero net moments in the directions as shown in the figures, so none of them are likely in truly antiferromagnetic states without any net magnetic moments. These non-zero net moments are likely due to orbital moments resulting from Berry curvature [40, 41]. When non-zero net moments exist, then the associated odd-order AHE will be dominated by a linear effect with current. However, any antiferromagnetic materials with 4', $\bar{4}$', 4'/$m$, 4'2'2 (4'22'), 4'$m$'$m$ (4'$mm$'), $\bar{4}$'$m$'2 ($\bar{4}$'2$m$'), $\bar{4}$'2'$m$ ($\bar{4}$'2$m$'), and 4'/$mm$'$m$ (4'/$mmm$') can exhibit odd-order AHE without any net magnetic moment. All MPGs described above have unbroken 4' or $\bar{4}$', so the odd-order AHEs in these point groups are associated with symmetric tensors, unlike antisymmetric tensors for odd-order AHEs in ferromagnetic space groups. For example, for $\bar{4}$' point group, (+***J$_x$***,+***E$_y$***) becomes (+***J$_y$***,+***E$_x$***) under the symmetry operation of **C$_{4z}$⊗I⊗T**, while the point group is invariant, so the relevant conductivity tensor components are symmetric (Figs. 3(h)&(i)). Odd-order AHE with symmetric tensors has never been reported and will be an exciting new research direction (see the section **IV** for discussions on specific compounds).

We emphasize that our SOS approach can tell if a certain phenomenon is zero, non-zero odd-order, or non-zero even-order effect, and broken {**I⊗T,T,M$_x$,M$_y$,R$_x$,R$_y$,C$_{3x}$,M$_z$⊗T,R$_z$⊗T**} is, in fact, the requirement for odd-order AHE, and those true antiferromagnetic point groups should exhibit high-odd-order, rather than linear, AHE. As we will discuss later, the requirement for odd-order AHE$_{yx}$ is identical with that for off-diagonal even-order current($x$) induced magnetization($z$); thus, current($x$) in those true antiferromagnets induces magnetization($z$) (for example, proportional to $J^2$) , which in turn, induces a Hall effect along $y$, so it becomes a high-odd-order (for example, proportional to $J^3$) effect. Recently, the concept of altermagnetism was introduced: their ordered spins are truly antiferromagnetic, but can exhibit, for example, non-zero linear AHE due to orbital magnetism through Berry curvature[34, 54, 55]. However, it turns out that all those altermagnets, showing linear AHE, Faraday effect, or MOKE effect discussed so far, belong to ferromagnetic point groups in terms of symmetry.

As we will discuss more later, current in crystallographically chiral systems can induce magnetization along the current direction as shown in Fig. 2(j)[56]. In some of these crystallographically chiral systems, magnetic texture can also have chirality, i.e., all mirror symmetries are broken even if any spatial rotations are freely allowed. "Magnetic toroidal moment + canted moment"[18], Bloch-type skyrmion and "magnetic quadrupole moment + alternating canted moment" in Fig. 3(e) are examples of these chiral magnetic textures in chiral crystallographic structures[28-30]. In these cases, current along $x$ induces magnetization along $x$, which, in turn induces voltage gradient (i.e., electric field or polarization) along $y$ axis, which is related with the origin of the so-called topological Hall effect for skyrmions[9-10]. It is interesting to note that all those chiral magnetic textures do have broken {**I**⊗**T**,**T**,**M**$_x$,**M**$_y$,**R**$_x$,**R**$_y$,**C**$_{3x}$,**M**$_z$⊗**T**,**R**$_z$⊗**T**}, so the topological Hall effect should exist in all those chiral spin textures in Fig. 3(e). Topological Hall effect for "magnetic toroidal moment + canted moment" or "magnetic quadrupole moment + alternating canted moment" has not been observed yet and needs to be confirmed. In fact, "magnetic toroidal moment + canted moment" was observed in BaCoSiO$_4$ (MPG 6) [57], "magnetic quadrupole + alternating canted moment" was observed in Er$_2$Ge$_2$O$_7$,[58] and Pb(TiO)Cu$_4$(PO$_4$)$_4$ (MPG 4'22') [59], and topological Hall effect in "magnetic quadrupole + alternating canted moment" can be one example of high-odd-order AHE in a true antiferromagnet without any net magnetic moment.

**B. Symmetry of piezomagnetism**

Piezoelectricity is the phenomena of inducing polarization, i.e., voltage gradient, with external stress. Similarly, piezomagnetism is the phenomena wherein a net magnetic moment (depicted as a blue arrow) is induced by external stress (represented by a green double arrow), as illustrated in Fig. 2(i). There can exist Diagonal or Off-diagonal piezomagnetism. The experimental setup for Diagonal piezomagnetism along the $x$-axis is 1D, which has unbroken [**1**,**I**,**M**$_x$,**R**$_x$,**M**$_y$⊗**T**,**M**$_z$⊗**T**,**R**$_y$⊗**T**,**R**$_z$⊗**T**] as shown in the equivalence relation from Fig. 2(i) ↔ 2(i). Conversely, in a reversed *M* condition the symmetry processes broken {**I**⊗**T**,**T**,**M**$_y$,**M**$_z$,**R**$_y$,**R**$_z$} with free rotation along the $x$-axis. Note that chirality is defined as lacking any mirror symmetry even any rotation around any axis is freely allowed. Here, we use the concept of free rotation around a given axis when we consider symmetry properties of one-dimensional objects or phenomena. This approach simplifies significantly the list of broken symmetries. For example, Diagonal piezomagnetism$_z$ requires {**I**⊗**T**,**T**,**M**$_x$,**M**$_y$,**R**$_x$,**R**$_y$} with free rotation along the $z$-axis, which is equal to broken {**I**⊗**T**,**T**,**M**$_x$,**M**$_y$,**R**$_x$,**R**$_y$,**R**$_z$,**R**$_z$**T**,**M**$_z$**T**,**C**$_{4z}$,**C**$_{4z}$⊗**T**,**C**$_{4z}$⊗**I**⊗**T**,**C**$_{3z}$,**C**$_{3z}$⊗**T**,**C**$_{3z}$⊗**I**⊗**T**,**C**$_{6z}$, **C**$_{6z}$⊗**T**,**C**$_{6z}$⊗**I**⊗**T**}. The complete lists of Diagonal piezomagnetism are shown in [13], [15], and [17] in TABLE 1. Similarly, Off-diagonal piezomagnetism with stress along $x$ and induced magnetization along $z$ has unbroken [**1**,**I**,**M**$_z$,**R**$_z$,**M**$_x$⊗**T**,**M**$_y$⊗**T**,**R**$_x$⊗**T**,**R**$_y$⊗**T**] and broken {**I**⊗**T**,**T**,**M**$_x$,**M**$_y$,**R**$_x$,**R**$_y$,**C**$_{3x}$,**M**$_z$⊗**T**,**R**$_z$⊗**T**}.

First, note that ferromagnetic point groups with magnetization along *x* has broken {**I**⊗**T**,**T**,**M**$_y$,**M**$_z$,**R**$_y$,**R**$_z$,**C**$_{3y}$,**C**$_{3z}$} with any free rotation along *x*, so all ferromagnetic point groups can exhibit Diagonal piezomagnetism along their magnetization direction as well as Off-diagonal piezomagnetism with induced magnetization along their initial magnetization direction.

Second, the requirement of {**I**⊗**T**,**T**,**M**$_x$,**M**$_y$,**R**$_x$,**R**$_y$,**C**$_{3x}$,**M**$_z$⊗**T**,**R**$_z$⊗**T**} for Odd-order AHE$_{yx}$ is identical with that for Off-diagonal piezomagnetism with stress along *x* and induced magnetization along *z*, so all systems exhibiting Odd-order AHE can show Off-diagonal piezomagnetism[60]. Third, in any piezomagnets, electric fields can also induce magnetization, but this electromagnetic effect will be even-order, i.e., the sign of electric fields does not matter. CoF$_2$ is a known piezomagnetic compound, and its magnetic point group is 4'/*mm*'*m* [53, 61, 62] with unbroken [**R**$_z$,**R**$_{xy}$,**R**$_{yx}$,**I**,**M**$_z$,**M**$_{xy}$,**M**$_{yx}$,**R**$_x$⊗**T**,**R**$_y$⊗**T**,**C**$_{4z}$⊗**T**,**M**$_x$⊗**T**,**M**$_y$⊗**T**,**C**$_{4z}$⊗**I**⊗**T**], so it can exhibit Off-diagonal piezomagnetism with stress along *x* or *y* and induced magnetization along *z*. Surprisingly, CoF$_2$, in fact, belongs to the circle for Odd-order AHE in Fig. 1, and can exhibit odd-order AHE$_{yx}$ and AHE$_{xy}$. Since it is insulating, CoF$_2$ can show non-zero anomalous Nernst and thermal Hall effects, which needs to be confirmed experimentally.

**C. Symmetry of the Even-order AHE measurements**

Now, we consider the experimental setup for even-order AHE with current along *x* and Hall voltage along *y* (Even-order AHE$_{yx}$) in Figs. 2(a). (a) ↔ (a) by **M**$_z$, **M**$_x$⊗**T**, and **R**$_y$⊗**T**; (a) ↔ (c) by **T**, **M**$_x$, **R**$_y$, and **M**$_z$⊗**T**; (a) ↔ (b) by **I**⊗**T**, **M**$_y$, **R**$_x$, and **R**$_z$⊗**T**; (a) ↔ (d) by **I**, **R**$_z$, **M**$_y$⊗**T**, and **R**$_x$⊗**T**. The experimental setup to measure even-order AHE$_{yx}$ has unbroken [**1**,**T**,**M**$_x$,**M**$_z$,**R**$_y$,**M**$_x$⊗**T**,**M**$_z$⊗**T**,**R**$_y$⊗**T**] from Fig. 2(a) ↔ (a) and (a) ↔ (c) and broken {**I**⊗**T**,**I**,**M**$_y$,**R**$_x$,**R**$_z$,**C**$_{3x}$,**M**$_y$⊗**T**,**R**$_x$⊗**T**,**R**$_z$⊗**T**} from Fig. Fig. 2(a) ↔ (b) and (a) ↔ (d). Any specimens, having SOS with this experimental setup, will show even-order AHE$_{yx}$ when they have broken {**I**⊗**T**,**I**,**M**$_y$,**R**$_x$,**R**$_z$,**C**$_{3x}$,**M**$_y$⊗**T**,**R**$_x$⊗**T**,**R**$_z$⊗**T**}([4] in TABLE 1). Note that Even-order is in terms of current (**J**), i.e., goes like V$_2$**J**$^2$ + V$_4$**J**$^4$ + - -- (V$_i$'s are constants). The unbroken [**1**,**T**,**M**$_x$,**M**$_z$,**R**$_y$,**M**$_x$⊗**T**,**M**$_z$⊗**T**,**R**$_y$⊗**T**] corresponds to MPG of *m*2*m*1'.

Monolayer H-MoS$_2$ ($\bar{6}$*m*21') [51], exhibiting quadratic AHE, does have broken {**I**⊗**T**,**I**,**M**$_y$,**R**$_x$,**R**$_z$,**C**$_{3x}$,**M**$_y$⊗**T**,**R**$_x$⊗**T**,**R**$_z$⊗**T**}. All polar groups do exhibit polarization, and do have broken {**I**⊗**T**,**I**,**R**$_x$,**R**$_y$,**C**$_{3x}$,**C**$_{3y}$,**R**$_x$⊗**T**,**R**$_y$⊗**T**} with free rotation along *z* when polarization is along *z*, so can exhibit Even-order AHE$_{zx}$ and AHE$_{zy}$ (Fig. 1, [2], and [6] in TABLE 1). There are a number of non-centrosymmetric and non-polar point groups having broken {**I**⊗**T**,**I**,**M**$_y$,**R**$_x$,**R**$_z$,**C**$_{3x}$,**M**$_y$⊗**T**,**R**$_x$⊗**T**,**R**$_z$⊗**T**} by setting current along *x* and Hall voltage along *y*, including 32, 32', 321', $\bar{4}$, $\bar{4}$1', $\bar{4}$', $\bar{4}$*m*2, $\bar{4}$*m*21', $\bar{4}$'*m*2', $\bar{4}$'*m*'2, $\bar{6}$, $\bar{6}$1', $\bar{6}$', $\bar{6}$*m*2, $\bar{6}$*m*21', $\bar{6}$*m*2', $\bar{6}$'*m*'2, and $\bar{6}$'*m*2' (Red and purple point groups in Fig. 1). Table 1 lists the correct orientations for Odd-order AHE and Even-order AHE of those MPGs. The MPGs of Even-order AHE derived from the tensor approach reported recently [31] are consistent with our symmetry methods except $\bar{6}$1' and $\bar{6}$*m*21'. $\bar{6}$1' and $\bar{6}$*m*21' can have high-even-order in terms of symmetry, and were proposed as 4$^{th}$ order induced by Berry curvature multipoles [63].

**D. Symmetry of piezoelectricity**

Piezoelectricity is often considered for non-magnetic states or point groups but can be extended to magnetic states or magnetic point groups. There can exist Diagonal, Off-diagonal, or shear-type piezoelectricity. The experimental setup for Diagonal piezoelectricity with stress and induced polarization along *x*, which is 1D, has unbroken [**1,T,M$_y$,M$_z$**] and broken {**I⊗T,I,R$_y$,R$_z$,R$_y$⊗T,R$_z$⊗T**} with free rotation along the *x*-axis. Off-diagonal piezoelectricity with stress along *x* and induced polarization along *y* (Off-diagonal piezoelectricity$_{yx}$) has unbroken [**1,T,M$_x$,M$_z$,R$_y$,M$_y$⊗T,M$_z$⊗T,R$_y$⊗T**] and broken {**I⊗T,I,M$_y$,R$_x$,R$_z$,C$_{3x}$,M$_y$⊗T,R$_x$⊗T,R$_z$⊗T**}, which is identical with the requirement for Even-order AHE$_{yx}$ or Off-diagonal odd-order current$_x$-induced ***M$_z$***. Thus, all MPGs in the Even-order AHE$_{yx}$ circle of Fig. 1 exhibit both Off-diagonal odd-order current$_x$-induced ***M$_z$*** and Off-diagonal piezoelectricity$_{yx}$.

Shear stress along *x* can be associated with a ferro-rotation axial vector along *x*, and the combination of ferro-rotation axial vector along *x* and polarization along *x* is identical with the combination of velocity vector along *x* and magnetization along *x* in terms of symmetry[21]. Therefore, all MPGs in the circle of Diagonal current-induced ***M*** exhibit shear stress-induced polarization. This shear stress-induced polarization usually occurs in chiral systems, but, for example, non-chiral $\bar{4}$1' can have nonzero d$_{14}$, i.e., polarization along *x* induced by shear stress along *x* [33]. More precisely speaking, shear stress along *z* changes its sign with **C$_{4z}$**, so, for example, MPG 4 with unbroken **C$_{4z}$** has zero d$_{36}$, i.e., zero polarization along *z* induced by shear stress along the *z*-axis ([33-36] in TABLE 1).

**E**. **Conjugated phenomena**

Opechowski [64] first brought attention to the symmetry analogies among magnetic, electric, and toroidal orderings within magnetic point groups, revealing the existence of "magic numbers," in the sense that each of ferromagnetic, ferroelectric, and ferrotoroidal orderings accommodates its own 31 magnetic crystal classes. Recent research[65] re-identified 122 magnetic point groups by considering the permutation features of magnetic (***M***), electric (***P***), and toroidal (***T***) modes, signifying broader symmetry connections among these ordering types. In this context, Magnetic (***M***) modes convert to electric (***P***) modes through **T↔I** transformation, while magnetic (***M***) modes are also shown to be convertible into toroidal (***T***) modes through **T↔I⊗T** transformation. Our work expands these connections, uncovering the interchangeable symmetry criteria between Odd-order AHE and Even-order AHE through the **T↔I** symmetry transformation as listed in Table 1 [1-12]. To illustrate, Odd-order AHE$_{yx}$ requires the breaking of the following symmetries {**I⊗T,T,M$_x$,M$_y$,R$_x$,R$_y$,C$_{3x}$,M$_z$⊗T,R$_z$⊗T**}, which can subsequently be transformed into the symmetry requirement for Even-order AHE$_{zx}$ {**I⊗T,I, R$_x$⊗T,R$_y$⊗T,R$_x$,R$_y$,C$_{3x}$,M$_z$⊗T,M$_z$**} via the **T↔I** symmetry transformation. Note that **T↔I** accompanies, for example, **M$_x$=R$_x$⊗I ↔ R$_x$⊗T** and **R$_z$⊗T↔ R$_z$⊗I=M$_z$**. Thus, Odd-order AHE and Even-order AHE can be exchanged with each other and may be regarded as "conjugate" phenomena. Thirty-one ferromagnetic MPGs (depicted as black and red MPGs in Fig. 1), which allow Odd-order AHE are interconnected with 31 polar MPGs (shown as black and blue MPGs in Fig. 1) that permit Even-order AHE. In detail, the number of MPGs, which allow for polar order along *x*, *y*, or *z*-axes is the same as the number of MPGs, which allow for ferromagnetic order along the same axes. Additionally, the remaining 24 non-ferromagnetic MPGs (represented by blue and purple MPGs in Fig. 1) conducive to Odd-order AHE are linked to 24 non-polar MPGs (depicted as red and purple MPGs in Fig. 1) that facilitate Even-order AHE. These connections exhibit intriguing consistency, as demonstrated in Table 1 [1-12]. Similarly, the **T↔I** symmetry transformation also establishes diagonal piezoelectricity and diagonal piezomagnetism as conjugate phenomena, as shown in Table 1 [13-18]. Note that the transformation between **T↔I** of chiral MPGs, which permits diagonal current-induced ***M*** (right-bottom circle of Fig. 1) preserves the chiral groups 222, 422, 622, 432, while the MPGs 2221', 4221', 6221', and 4321' are transformed into *mmm*, 4/*mmm*, 6/*mmm*, and *m$\bar{3}$m*, respectively. These transformed MPGs don't exhibit phenomena discussed along principal axes, i.e., they fall outside three circles.

**F. Symmetry of the measurements for off-diagonal current-induced magnetization *M***

We now consider the experimental setup for odd-order (or linear) off-diagonal current-induced magnetization *M* by current (Fig. 2(e)-(h)), whose transformation properties are like this: (e) ↔ (e) by **1**, **M$_z$**, **M$_x$**⊗**T**, and **R$_y$**⊗**T**; (e) ↔ (g) by **T**, **M$_x$**, **R$_y$**, and **M$_z$**⊗**T**; (e) ↔ (f) by **I**⊗**T**, **M$_y$**, **R$_x$**, **C$_{3x}$** and **R$_z$**⊗**T**; (e) ↔ (h) by **I**, **R$_z$**, **R$_x$**⊗**T** and **M$_y$**⊗**T**. Thus, the setup has unbroken [**1**,**T**,**M$_x$**,**M$_z$**,**R$_y$**,**M$_x$**⊗**T**,**M$_z$**⊗**T**,**R$_y$**⊗**T**] from Fig. 2(e) ↔ (e) and 2(e) ↔ (g) and broken {**I**⊗**T**,**I**,**M$_y$**,**R$_x$**,**R$_z$**,**C$_{3x}$**,**M$_y$**⊗**T**, **R$_x$**⊗**T**,**R$_z$**⊗**T**} from Fig. 2(e) ↔ (f) and 2(e) ↔ (h). In fact, these are identical with the ones for the experimental setup for Even-order AHE$_{yx}$. Thus, Even-order Hall effect works basically like this: current induces magnetization or magnetic field along a 2$^{nd}$ direction, then Hall voltage develops along the 3$^{rd}$ direction, and the induced magnetization varies linearly (or in an odd-order) with current, expressed as *M*⊥($k^{2n+1}$) in Fig. 1, so the induced Hall voltage changes like square (or even-order) of current.

Bulk T$_d$-WTe$_2$ (*Pmn*2$_1$) has unbroken (**R$_z$**,**M$_x$**,**M$_y$**), but **R$_z$** and **M$_y$** are broken bilayer T$_d$-WTe$_2$ (point group *m*), [66] so bilayer T$_d$-WTe$_2$ has broken {**I**⊗**T**,**I**,**M$_y$**,**R$_x$**,**R$_z$**,**C$_{3x}$**,**M$_y$**⊗**T**,**R$_x$**⊗**T**,**R$_z$**⊗**T**}, and, thus, shows Off-diagonal linear current$_x$-induced *M$_z$*, which requires the experimental confirmation. In addition, bilayer T$_d$-WTe$_2$ does show quadratic AHE with Hall voltage along *y*. [49] Note that monolayer T$_d$-WTe$_2$ (*P*2$_1$/*m*) has unbroken (**R$_x$**,**I**,**M$_x$**), so cannot exhibit neither linear current-induced magnetization nor nonlinear AHE. The bilayer system discussed for the Edelstein effect has also broken {**I**⊗**T**,**I**,**M$_y$**,**R$_x$**,**R$_z$**,**C$_{3x}$**,**M$_y$**⊗**T**,**R$_x$**⊗**T**,**R$_z$**⊗**T**}, so can exhibit Off-diagonal linear current$_x$-induced *M$_z$*. [67]

We have discussed current-induced effects, but emphasize that our symmetry argument works for any moving quasiparticles such as electric current, light, magnons, or phonons [68]. For example, monolayer H-MoS$_2$ or *h*-BN ($\bar{6}$*m*21' in Fig. 3(k)) having broken {**I**⊗**T**,**I**,**M$_y$**,**R$_x$**,**R$_z$**,**C$_{3x}$**,**M$_y$**⊗**T**,**R$_x$**⊗**T**,**R$_z$**⊗**T**}, is supposed to show magnetization along the out-of-plane direction with phonons along a particular in-plane direction, which is sometimes called chiral phonons[13]. Note that these phonons do not have all broken mirror symmetries, so they are achiral, so the terminology of chiral phonons is not appropriate. All polar point groups and non-centrosymmetric and non-polar point groups (Red and purple point groups in Fig. 1) mentioned in section C can have Off-diagonal odd-order current-induced *M*. Note that $\bar{6}$1' and $\bar{6}$*m*21' don't show true-linear current-induced magnetization due to zero gyrotropic tensors[28]; however, off-diagonal high-odd-order (such as 3$^{rd}$-order) current-induced magnetization should be allowed in term of symmetry.

Let's now discuss Off-diagonal even-order current-induced magnetization. Consider the experimental setup for measuring magnetization along a direction perpendicular to the current direction induced quadratically (or in an even-order) by current (Fig. 2(e)-(h)), whose transformation properties are like this: (e) ↔ (e) by **1**, **M$_z$**, **M$_x$**⊗**T**, and **R$_y$**⊗**T**; (e) ↔ (h) by **I**, **R$_z$**, **R$_x$**⊗**T** and **M$_y$**⊗**T**; (e) ↔ (f) by **I**⊗**T**, **M$_y$**, **R$_x$**, **C$_{3x}$** and **R$_z$**⊗**T**; and (e) ↔ (g) by **T**, **M$_x$**, **R$_y$**, and **M$_z$**⊗**T**. Thus, the setup has unbroken [**1**,**I**,**M$_z$**,**R$_z$**,**M$_x$**⊗**T**,**M$_y$**⊗**T**,**R$_x$**⊗**T**,**R$_y$**⊗**T**] from Fig. 2(e) ↔ (e) and 2(e) ↔ (h) and broken {$I \otimes T, T, M_x, M_y, R_x, R_y, C_{3x}, M_z \otimes T, R_z \otimes T$} from Fig. 2(e) ↔ (f) and 2(e) ↔ (g). In fact, these are identical with the ones for the experimental setup for Odd-order AHE$_{yx}$. As we have discussed, High-odd-order AHE for true antiferromagnets such as 4'2'2 and 4'/$mm'm$ works basically like this: current induces magnetization quadratically (or in an even-order) along a 2$^{nd}$ direction, then Hall voltage develops along the 3$^{rd}$ direction, and the induced magnetization varies quadratically (or in an even-order) with current, expressed as $M\perp(k^{2n})$ in Fig. 1, so the induced Hall voltage changes like in a high-odd-order with current.

The requirements for Odd-order AHE, Even-order AHE and Off-diagonal current-induced $M$ are summarized in Table 1, and the relevant MPGs are classified in Fig. 1. Note that all MPGs discussed in conjunction with "crystal Hall effects" [34] are ferromagnetic point groups and a part of the circle for Odd-order AHE in Fig. 1. Table 1 and Fig. 1 also list the requirements and all MPGs for Diagonal odd-order current-induced magnetization and optical activity, respectively, which will be discussed below.

### G. Symmetry of the measurements for diagonal current-induced magnetization $M$

Electric current flow in chiral conductors can induce magnetization along the current direction, expressed as $M_{//}(k^{2n+1})$ in Fig. 1, and this phenomenon has been observed, for example, in tellurium crystals with a mono chiral domain state[56]. Fig. 2(k) depicts the experimental setup to measure this diagonal odd-order current-induced magnetization along $x$. Here, we disregard any rotations along $x$, since this experimental setup has invariant under any rotations along $x$. This setup has unbroken [$1, T, R_y, R_z, R_y \otimes T, R_z \otimes T$] as shown in the equivalence relation from Fig. 2(k) ↔ 2(k) and this holds true in a reversed $M$ and $J$ condition. However, under either a reversed $M$ or a reversed $J$ condition, the symmetry processes broken {$I \otimes T, I, M_y, M_z, M_y \otimes T, M_z \otimes T$} with free rotation along the $x$-axis. Thus, the diagonal odd-order current-induced $M$ requires broken {$I \otimes T, I, M_y, M_z, M_y \otimes T, M_z \otimes T$} with free rotation along the $x$-axis ([22-24] in TABLE 1). Note that one exemplary MPG with unbroken [$1, T, R_x, R_y, R_x \otimes T, R_y \otimes T$] with free rotation along $z$ is 2221'. All chiral point groups do satisfy these requirements.

There are eight non-chiral point groups where {$I \otimes T, I, M_y, M_z, M_y \otimes T, M_z \otimes T$} are all broken ($\bar{4}$, $\bar{4}1'$, and $\bar{4}'$ shown in Figs. 3(f)-(h) as well as $\bar{4}2m$, $\bar{4}2m1'$, $\bar{4}2'm'$, $\bar{4}'2'm$, and $\bar{4}'2m'$), so should exhibit Diagonal current-induced magnetization. For $\bar{4}1'$, current along $x$ ($y$) can induce $M_x$ ($M_y$) and Even-order AHE voltage along $z$. For both $\bar{4}$ and $\bar{4}'$ current along $x$ ($y$) can induce Odd-order AHE voltage along $y$ ($x$) and Even-order AHE voltage along $z$ (so $M_y$ ($M_x$)). In addition, current along $x$ ($y$) in $\bar{4}$ and $\bar{4}'$ can also induce $M_x$ ($M_y$) (see Figs. 3(h) & (i)). Current along $x$ and induced Hall voltage along $y$ can induce magnetization along $z$, and this induced $M$ is quadratic (or in an even-order) with current, so e.g., the switching of current direction does not change the direction of induced $M_z$. Note that many ferromagnetic point groups with current along ferromagnetic moment has broken {$I \otimes T, I, M_y, M_z, M_y \otimes T, M_z \otimes T$}, so can exhibit diagonal current-induced magnetization along the ferromagnetic moment; however, this induced magnetization may be significantly smaller than the preexisting ferromagnetic moment. Thus, we did not include these ferromagnetic point groups for the diagonal current-induced $M$ in the Fig. 1 diagram.

One interesting phenomenon associated with the diagonal odd-order current-induced magnetization is the control of the chirality of helical spin states using the simultaneous presence of current and magnetic fields: helical states do have broken {$I \otimes T, I, M_y, M_z, M_y \otimes T, M_z \otimes T$}, and, in fact, are chiral since all mirrors symmetries are broken without time reversal breaking even if any spatial rotations are allowed, so current along $x$ can induce $M_x$ or -$M_x$, depending on the chirality, so the relative sign of coexisting current and magnetic field favors energetically one chirality vs the other chirality, so can control the chirality of helical spin states [69]. Each Bloch-type skyrmion or magnetic bubble is chiral since all mirror symmetries are broken without time reversal breaking even if any spatial rotations are allowed. Note that the Bloch-type skyrmion in Fig. 3(l) corresponds to 42'2' point group, which is chiral, ferromagnetic, and non-polar. Bloch-type skyrmions have mono chirality and magnetic bubbles have random chirality. From the above consideration, we predict this: magnetic bubbles with mono chirality can be induced using the simultaneous presence of current and magnetic field along one direction, and the change of the relative sign of coexisting current and magnetic field can induce the chirality flipping.

Note that the Diagonal current-induced $M$ discussed above is odd-order (likely linear) effects, but we can also consider Diagonal even-order current-induced $M$. The requirement for Diagonal even-order current-induced $M$ along $z$ is broken {$I \otimes T, T, M_x, M_y, R_x, R_y$} with any free rotation along $z$, which is precisely aligned with Diagonal piezomagnetism$_z$ ([13], [15], and [17] in TABLE 1). The requirement is similar with the requirement of broken {$I \otimes T, T, M_x, M_y, R_x, R_y, C_{3x}, C_{3y}$} for a net magnetic moment along $z$ – the only difference is $C_{3x}$ and $C_{3y}$. Thus, all MPGs belonging to the ferromagnetic point group exhibit Diagonal even-order current-induced $M$. In addition, these MPGs with $C_{3z}$ can exhibit Diagonal even-order current-induced $M_x$ or $M_y$: 3, $\bar{3}$, 32, 3$m$, $\bar{3}m$, 6', $\bar{6}$', 6'/$m$', 6'22', 6'$mm$', $\bar{6}$'$m$2', $\bar{6}$'2$m$', 6'/$m$'$mm$' along $x$ and 3, $\bar{3}$, 32', 3$m$', $\bar{3}m$', 6', $\bar{6}$', 6'/$m$', 6'$m$'$m$, 6'2'2, $\bar{6}$'2'$m$, $\bar{6}$'$m$'2, 6'/$m$'$m$'$m$ along $y$. Note that Diagonal even-order current-induced $M$ is analogous to magnetostriction effects.

## H. Symmetry of Directional nonreciprocity in $H$

When the motion of an object in a specimen in one direction is different from that in the opposite direction, it is called a nonreciprocal directional dichroism or simply a directional nonreciprocity. The object can be an electron, a phonon, spin wave, light in crystalline solids, or the specimen itself, and the best-known example is that of nonreciprocal charge transport (i.e., diode) effects in $p$–$n$ junctions, where a built-in electric field ($E$) breaks the directional symmetry. In addition to $p$–$n$ junctions, numerous technological devices, such as optical isolators, spin current diodes, or metamaterials do utilize nonreciprocal effects. Nonreciprocity directional dichroism can often be understood from the consideration of SOS with velocity vector, $k$. Herein, we focus on Directional nonreciprocity in the presence of external magnetic field $H$. We first consider the $H_x$//$k_x$ case, i.e., Directional nonreciprocity$_x$ in $H_x$, which can be represented by Fig. 2(k), and is the same situation with Diagonal odd-order (linear) current-induced $M_x$. Thus, broken {$I \otimes T, I, M_y, M_z, M_y \otimes T, M_z \otimes T$} with any free rotations along $x$ is the requirement for Directional nonreciprocity$_x$ in $H_x$, and is, in fact, also that for Diagonal current-induced $M_x$. Thus, all MPGs for non-zero Diagonal current-induced $M_x$ can exhibit non-zero Directional nonreciprocity$_x$ in $H_x$. ([22-24] in TABLE 1).

We can also consider the $H_z \perp k_x$ case, i.e., Directional nonreciprocity$_x$ in $H_z$, which can be represented by Fig. 2(e) and is similar with Off-diagonal odd-order current$_x$-induced $M_z$ or Even-order AHE$_{yx}$. Since applying $H_z$ can break $C_{3x}$, Directional nonreciprocity$_x$ in $H_z$ requires broken {$I \otimes T, I, M_y, R_x, R_z, M_y \otimes T, R_x \otimes T, R_z \otimes T$} even though Off-diagonal odd-order current$_x$-induced $M_z$ or Even-order AHE$_{yx}$ requires broken {$I \otimes T, I, M_y, R_x, R_z, C_{3x}, M_y \otimes T, R_x \otimes T, R_z \otimes T$}. Thus, all magnetic point groups for non-zero Off-diagonal odd-order current$_x$-induced $M_z$ or Even-order AHE$_{yx}$ can exhibit non-zero Directional nonreciprocity$_x$ in $H_z$ ([19-21] in TABLE 1).

Due to $C_{3z}$, 32, 321', and 32' exhibit zero Off-diagonal odd-order current$_z$-induced $M_y$ or Even-order AHE$_{xz}$ even though they can show non-zero Directional nonreciprocity$_z$ in $H_y$. Similarly, due to $C_{3z}$, $\bar{6}$, $\bar{6}1'$, and $\bar{6}'$ have zero Off-diagonal odd-order current$_z$-induced $M_y$ (or $M_x$) or Even-order AHE$_{xz}$ (or AHE$_{yz}$) even though they can show non-zero Directional nonreciprocity$_z$ in $H_y$ (or $H_x$). In addition, due to $C_{3z}$, $\bar{6}m2$, $\bar{6}m21'$, $\bar{6}'m'2$, $\bar{6}'m2'$, and $\bar{6}m'2'$ have zero Off-diagonal odd-order current$_z$-induced $M_x$ or Even-order AHE$_{yz}$ even though they can show non-zero Directional nonreciprocity$_z$ in $H_x$. Most of these cases are high-order effects and have not been explored neither theoretically nor experimentally, so can be a topic for the future investigation.

### I. Symmetry of the optical activity measurements

Experimental setup to measure optical activity is shown in Figs. 2(l) & (m). Optical activity includes circular dichroism[70], the Faraday effect[71], natural optical activity, and Circular PhotoGalvanic Effect (CPGE)[72, 73]. Circularly polarized light inherently carries spin angular momentum along its direction of propagation, endowing it with chiral characteristics. We depict circularly polarized light using (counter-)clockwise arrows atop wiggled arrows, symbolizing its chiral nature. Circularly polarized light, acting as a chiral entity, is represented in Figs. 2(l) with right-handed circularly polarized light and 2(m) with left-handed circularly polarized light, linked through {$\mathbf{I} \otimes \mathbf{T}, \mathbf{M}_x, \mathbf{M}_y$} with any spatial rotation along $z$. However, each setup is invariant under [$\mathbf{1}, \mathbf{R}_x \otimes \mathbf{T}, \mathbf{R}_y \otimes \mathbf{T}$] with any spatial rotation along $z$. Thus, broken {$\mathbf{I} \otimes \mathbf{T}, \mathbf{M}_x, \mathbf{M}_y$} with free rotation along $z$ is required to have optical activity$_z$. Emphasize that for this symmetry consideration, we allow any spatial rotation along $z$ freely ([25-27] in TABLE 1).

The broken-symmetry requirement for Optical activity is a subset of those for Odd-order AHE and Diagonal current-induced $M$, so most of MPGs for non-zero Odd-order AHE except 4'/$m$, 4'$m'm$, 4'/$mmm'$ $m\bar{3}$, 4'32', $\bar{4}'3m'$, and $m\bar{3}m'$ and Diagonal current-induced $M$ such as AgGaS$_2$ with MPGs $\bar{4}2m1'$ [74] belong to the class for optical activity. In the case of all ferromagnetic point groups, which is a part of ones for Odd-order AHE, Optical activity is always along the magnetization direction, and in the case of all chiral point groups, which is a part of ones for Diagonal current-induced $M$, Optical activity can be any of $x$, $y$, $z$ directions. Note that the optical activity tensor g$_{ij}$ of $\bar{3}m$, $3m$, $\bar{6}'$, $6'/m'$, $6'mm'$, $\bar{6}'m2$, $\bar{6}'m2'$ and $6'/m'mm'$ are zero due to the existence of three-fold C$_{3z}$ symmetry[33], however, we argument that the light induced C$_3$ breaking from experimental setup can give rise to Optical activity in those systems, likely as high-order effects.

### J. Coupling with $k//M$ and $k \perp M$

In terms of symmetry, there is no difference between spin angular momentum, orbital angular momentum, and magnetization. In addition, circularly-polarized light can be considered as light with $k//M$ (Fig. 2(n)), where $k$ is linear momentum of light and $M$ is spin angular momentum of light, so optical activity stems from non-zero coupling between light with $k//M$ and materials. A similar situation with $k//M$ can occur in polarized neutron scattering or spin-polarized scanning tunneling microscopy (SP-STM) experiments, so all materials with non-zero optical activity, i.e. those for non-zero Diagonal current-induced $M$, non-zero Odd-order AHE except 4'/$m$, 4'$m'm$, 4'/$mmm'$, $m\bar{3}$, 4'32', $\bar{4}'3m'$, and $m\bar{3}m'$ (Fig. 1) can exhibit non-trivial coupling in polarized neutron scattering or SP-STM experiments with $k//M$. However, for polarized neutron scattering or SP-STM experiments. $M$ can be perpendicular to $k$ (Fig. 2(o)), so many additional materials can exhibit non-zero activity in polarized neutron scattering or spin-polarized scanning tunneling (SP-STM) experiments. $k_x \perp M_z$ in Fig. 2(o) is linked through {$I \otimes T, M_y, R_x, R_z \otimes T, C_{3x}$} with a condition of $+k_x$ but a reversed $M_z$. Consequently, materials exhibiting non-zero activity in the $k_x \perp M_z$ experiments should have broken {$I \otimes T, M_y, R_x, R_z \otimes T, C_{3x}$}. Broken {$I \otimes T, M_y, R_x, R_z \otimes T, C_{3x}$} is a subset of broken {$I \otimes T, I, M_y, R_x, R_z, M_y \otimes T, R_x \otimes T, R_z \otimes T$} for Directional nonreciprocity$_x$ in $H_z$, which is again a subset of {$I \otimes T, I, M_y, R_x, R_z, C_{3x}, M_y \otimes T, R_x \otimes T, R_z \otimes T$} for Even-order AHE$_{yx}$. Therefore, all materials for non-zero Directional nonreciprocity$_x$ in $H_z$, including all materials for non-zero Even-order AHE$_{yx}$, can exhibit non-zero activity in the $k_x \perp M_z$ experiments. In addition, all MPGs inside the Odd-order AHE circle (Fig. 1) can exhibit non-trivial coupling in polarized neutron scattering or SP-STM experiments with $k \perp M$ ([28-33] in TABLE 1).

## IV. CANDIDATE MATERIALS
(A) Hexagonal crystal family

Based on Fig. 1, one can readily find new materials to measure various steady-state phenomena that we have discussed. 16 magnetic point groups inside the middle of the diagram of Fig. 1 can exhibit all phenomena discussed. For example, metallic cubic Pd$_3$Mn [75] and insulating NaMnFeF$_6$ [76] form in ferromagnetic, chiral and non-polar 32′ with unbroken ($C_{3z}, R_x \otimes T$). The 32' MPG, allowing all of these phenomena, e.g., Odd-order AHE$_{yx}$, Odd-order AHE$_{xy}$, Even-order AHE$_{xy}$, Diagonal current-induced $M_x$, $M_y$, Optical activity$_{x, y \text{ or } z}$, Diagonal piezomagnetism$_{y \text{ and } z}$ and Off-diagonal piezomagnetism can occur in Pd$_3$Mn and NaMnFeF$_6$. Insulating NaMnFeF$_6$ can show Odd-order or Even-order thermal AHEs, Diagonal light-induced $M$, and Optical activities. BaCoSiO$_4$ forms in the ferromagnetic, polar and chiral point group of 6 with unbroken ($C_{6z}, C_{3z}, C_{2i}$), and can exhibit Odd-order AHE$_{yx}$, Even-order AHE$_{zy \text{ or } zx}$, Diagonal current-induced $M_{x, y, \text{ or } z}$, Optical activity$_{x, y \text{ or } z}$ and Off-diagonal piezomagnetism. Chiral, non-ferromagnetic, and non-polar 321′ with unbroken [$C_{3z}, R_x, T, C_{3z} \otimes T, R_x \otimes T$] locates inside a two-circle intersection, and metallic SrFeO$_{3-\delta}$ [77] and ScFeGe [78] form in 321′. Thus, SrFeO$_3$ and ScFeGe can exhibit Even-order AHE$_{xy}$ and Diagonal current-induced $M_x$, $M_y$, or $M_z$ and Optical activity$_{x, y, \text{ or } z}$, and Directional nonreciprocity$_z$ in $H_y$. Emphasize that the current can be electric current or other propagating quasiparticles such as thermal current, propagating light, magnons and phonons. MPG of CsFeCl$_3$ at 2GPa below T$_N$=4.7 K is $\bar{6}'m2'$, while the point group of it above T$_N$ is centrosymmetric 6/$mmm$[79]. $\bar{6}'m2'$ has unbroken [$C_{6z} \otimes I \otimes T, C_{3z}, M_x, R_y \otimes T, M_z \otimes T$], so can exhibit Diagonal piezoelectricity$_x$, without any polarization in zero stress, which is solely due to magnetic order.

(B) Tetragonal crystal family

Insulating CsCoF$_4$ forms in $\bar{4}'$ [80], so it can exhibit Odd-order thermal AHE$_{yx}$, Odd-order thermal AHE$_{xy}$, Even-order thermal AHE$_{zy}$, Even-order thermal AHE$_{zx}$, Diagonal thermal-current-induced $M_x$, $M_y$, Optical activity$_{x, y \text{ or } z}$, and Off-diagonal piezomagnetism. Besides, as we discussed earlier, Er$_2$Ge$_2$O$_7$, and Pb(TiO)Cu$_4$(PO$_4$)$_4$ forms in chiral, non-polar and non-ferromagnetic 4'22' with unbroken [$C_{4z} \otimes T, R_z, R_x, R_y, R_{xy} \otimes T, R_{yx} \otimes T$] (here, the $x$ and $y$ directions are in accordance with 4'22', rather than those in Fig. 3(l)), so Odd-order AHE$_{yx,xy}$, Diagonal current-induced $M_{x, y, \text{ or } z}$, Off-diagonal piezomagnetism, and Optical activity$_{x, y \text{ or } z}$ can occur in Pb(TiO)Cu$_4$(PO$_4$)$_4$ and Er$_2$Ge$_2$O$_7$. We emphasize that this Odd-order AHE can be observed as an anomalous thermal Hall effect in those insulating 4'22' systems without any ferromagnetic moment, as shown in Fig. 3(l).

## V. DISCUSSION AND CONCLUSIONS

Using the SOS approach, we have identified the requirements and the relevant magnetic point groups for Odd-order AHE, Even-order AHE, Diagonal piezomagnetism, Off-diagonal piezomagnetism, Diagonal current-induced magnetization, Off-diagonal odd-order current-induced magnetization, Off-diagonal even-order current-induced magnetization, Directional nonreciprocity in a transverse magnetic field, and optical activity. Typically, Odd-order AHE, Even-order AHE and Diagonal linear current-induced magnetization occur in ferromagnetic, polar and chiral point groups, respectively. However, we have identified several point groups, which do not belong to those ferromagnetic/polar/chiral point groups but can exhibit those phenomena. Even-order AHE and Off-diagonal odd-order current-induced magnetization are basically the identical phenomenon, and High-odd-order AHE in true antiferromagnets is to Off-diagonal even-order current-induced magnetization. Off-diagonal piezomagnetism with stress along $x$ and induced magnetization along $z$, is identical with Odd-order AHE$_{yx}$, so all systems exhibiting Odd-order AHE can show Off-diagonal piezomagnetism.

Magnetization induced by electric current can be relevant to numerous subsequent phenomena, in addition to inducing Odd-order or Even-order AHE. For example, $M_x$ induced by electric current $k_x$ in Fig. 2(n) can switch the antiferromagnetic state in Fig. 2(q) to that in Fig. 2(p) through a spin-flop transition. Similarly, $M_z$ induced by electric current $k_x$ in Fig. 2o can switch the antiferromagnetic state in Fig. 2(p) to that in Fig. 2(q) through a spin-flop transition. Another example is a directional non-reciprocal effect in monolayer H-MoS$_2$ ($\bar{6}m21'$): electric current along $x$ induces $M_z$, so there exists a directional non-reciprocal effect of electric current when $H_z$ is present in additon to electric current in monolayer H-MoS$_2$. These experiments have not been performed and need to be verified in the future.

The discussed phenomena with $k$ can occur not only with electric current, but also with other moving quasi-particles such as light, phonons, and magnons. Therefore, for example, phonons in any Odd-order AHE systems can accompany induced Even-order magnetization that is perpendicular to the phonon $k$. Similarly, phonons in any Even-order AHE systems can accompany induced Odd-order magnetization that is perpendicular to the phonon $k$. Furthermore, phonons in any Diagonal current-induced $M$ systems can accompany induced magnetization that is along the phonon $k$. All of these phonons with any orientation of induced magnetization can be referred to as "magnetic phonons" in general, and phonons with induced magnetization along the phonon $k$ in any Diagonal current-induced $M$ systems can be referred to as "chiral phonons", which is a subset of magnetic phonons. For example, phonons in monolayer WSe$_2$ ($\bar{6}m21'$, belonging to the Even-order AHE MPGs) can accompany induced perpendicular magnetization, so become magnetic phonons [13]. Phonons in chiral systems such as α-HgS [81] and quartz [82] naturally accompany induced parallel magnetization, so become chiral phonons (and magnetic phonons). Emphasize that all phonons with non-zero $k$ in chiral systems must be chiral. Non-chiral MPGs such as $\bar{4}$, $\bar{4}1'$, $\bar{4}'$, $\bar{4}2m$, $\bar{4}2m1'$, $\bar{4}2'm'$, $\bar{4}'2'm$, $\bar{4}'2m'$ (Dashed-line box of Fig. 1) can accompany chiral phonons, since they allow Diagonal current-induced $M$. Among non-chiral MPGs showing Diagonal current-induced $M$, only $\bar{4}1'$ and $\bar{4}2m1'$ are non-magnetic. Thus, "chiral phonons in non-chiral and non-magnetic systems" are expected in compounds in paramagnetic/non-magnetic states with $\bar{4}1'$ (e.g., BPO$_4$, LiBSiO$_4$ and InPS$_4$)[83-85] or $\bar{4}2m1'$ (e.g., Ca$_2$MgSi$_2$O$_7$, BaS$_3$ and AgGaS$_2$)[74, 86-87].

Note also that AHE is not just for the standard anomalous Hall effect, but includes anomalous Ettingshausen, anomalous Nernst, and anomalous thermal Hall effects. We stress that our SOS approach does not consider specific coupling terms, so different from the standard tensorial approach, and can tell if an effect is zero, non-zero odd-order or non-zero even-order. Finally, we would like to emphasize that our results based on symmetry considerations are far-reaching and rather universal; however, the estimation of the magnitude of observable effects cannot be done only from symmetry considerations and requires the understanding of microscopic mechanisms. Undoubtedly, our comprehensive results will motivate future extensive investigations theoretically as well as experimentally. Clearly, our findings will be an essential guidance in discovering new non-traditional materials exhibiting those discoursed phenomena, and in general, the SOS approach can be utilized to unveil the requirements and relevant point groups for numerous other complex phenomena, which have not been explored in traditional symmetry approaches.

**ACKNOLEDGEMENT:** We have great benefits from discussions with David Vanderbilt, Sobhit Singh, Seong Joon Lim, Kai Du, Hanyu Zu, and Choongjae Won. This work was supported by the center for Quantum Materials Synthesis (cQMS), funded by the Gordon and Betty Moore Foundation's EPiQS initiative through grant GBMF10104, and by Rutgers University.

**COMPETING INTERESTS:** The authors declare no competing interests.
**DATA AVAILABILITY:** All study data is included in the article.

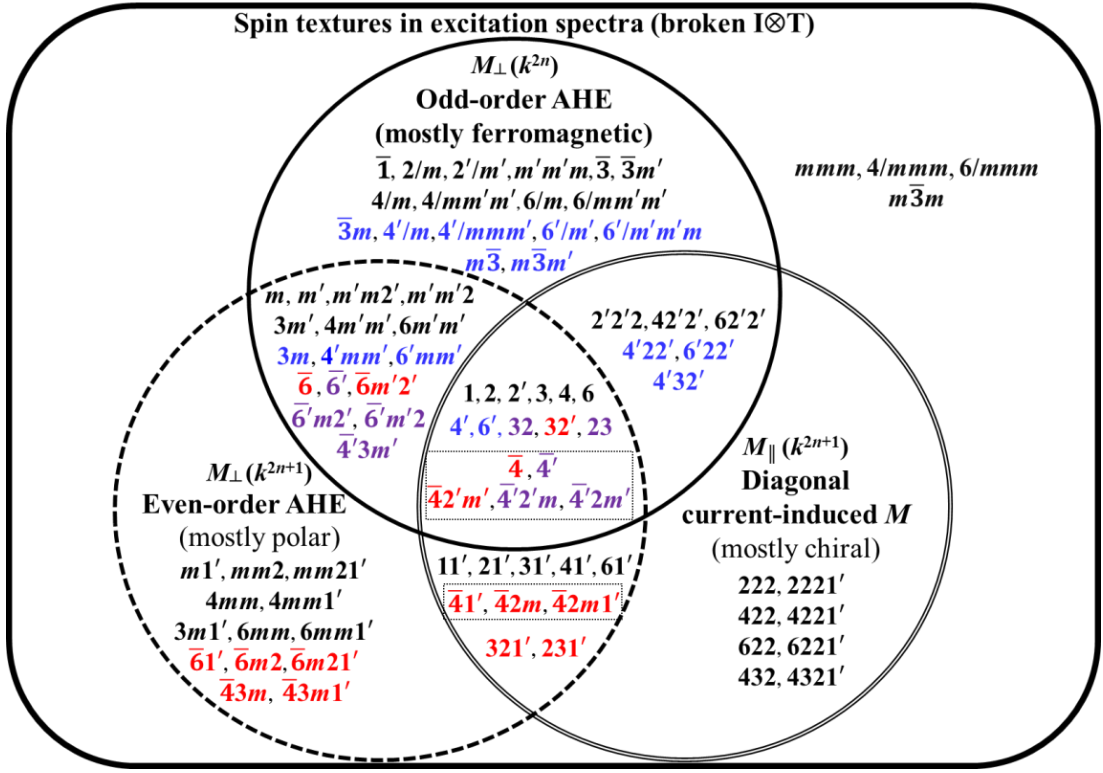

Fig. 1. Kinetomagnetism of magnetic point groups (MPGs) with broken **I**⊗**T**. This trinity diagram represents how magnetization (*M*), is induced by (electric) currents (*k*) in various MPGs with broken **I**⊗**T**. Top solid circle: Odd-order AHE, Off-diagonal piezomagnetism, and Even-order Off-diagonal current-induced magnetization (*M*), denoted as $M\perp(k^{2n})$, where $n$ is an integer. Left dashed circle: Even-order AHE, Off-diagonal piezoelectricity, and Odd-order Off-diagonal current-induced *M*, expressed as $M\perp(k^{2n+1})$. Right double-line circle: Diagonal Odd-order current-induced *M*, noted as $M//(k^{2n+1})$, and Diagonal directional nonreciprocity in *H*. Blue: non-ferromagnetic; Red: non-polar; Purple: non-ferromagnetic and non-polar. Dashed-line box: non-chiral. $\bar{4}'$ is non-polar, non-ferromagnetic, and non-chiral as shown in purple and in dashed-line box. All MPGs showing either Odd-order AHE (solid circle) or Diagonal Odd-order current-induced *M* (double-line circle), except 4'/*m*, 4'*mm*', 4'/*mmm*', $m\bar{3}$, $\bar{4}$'3*m*', and $m\bar{3}m$' can show optical activity and *k*//*M* coupling. *k*⊥*M* coupling exists in either Odd-order AHE (solid circle) or Even-order AHE (dashed circle). Four MPGs (*mmm*, 4/*mmm*, 6/*mmm*, and $m\bar{3}m$), even though they do have broken **I**⊗**T**, do not belong to any of the three circles and do not exhibit any phenomena in directions that considered, but they can show magnetic textures in excitation spectra. For symmetry consideration, we allow freely any spatial rotations along the 1D direction for the 1D phenomena of optical activity and diagonal Odd-order current-induced *M*.

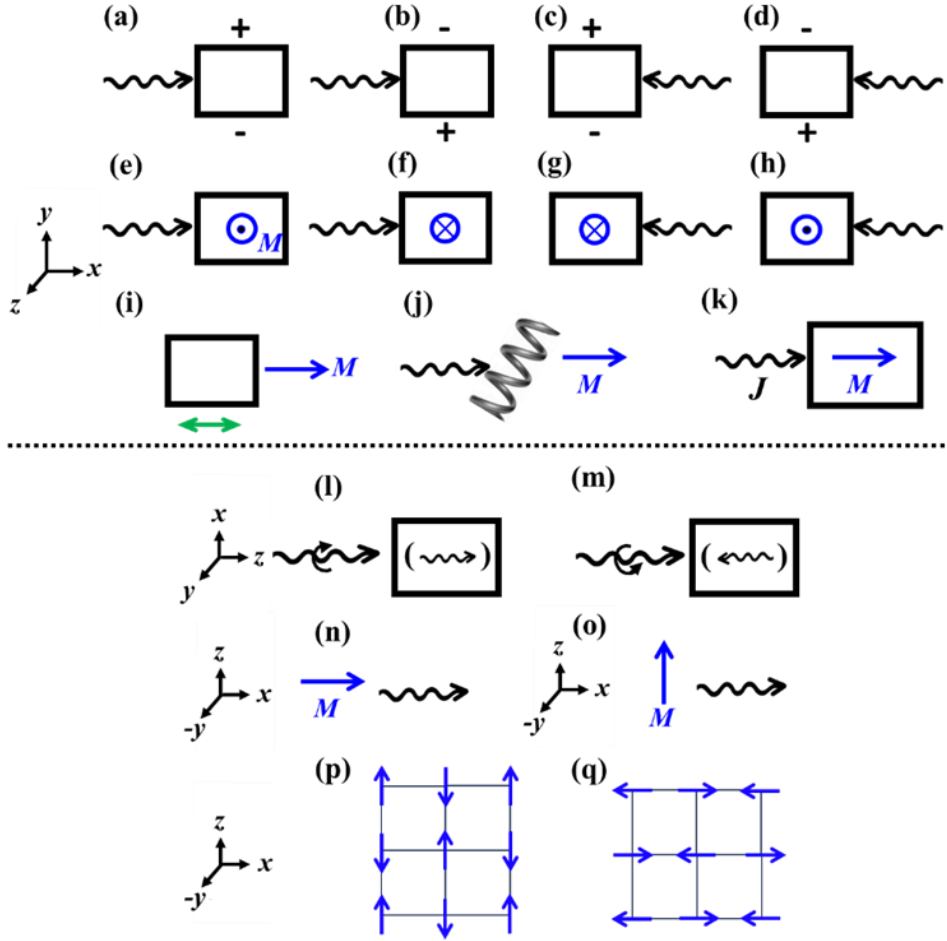

Fig. 2. Various experimental setups. (a)-(d) Four experimental setups to measure AHE. Wiggled arrows denote applied current. (e)-(h) Four experimental setups to measure magnetization, perpendicular to the current direction, induced by current. (i) Diagonal piezomagnetism. A green double arrow denotes a stress force. (j) Diagonal current-induced magnetization in chiral systems. (k) Experimental setup to measure magnetization along $x$ induced by current along $x$. (l) & (m) opposite circularly-polarized light illuminations. The small, wiggled arrows in parathesis depict electric currents induced by circularly-polarized light illumination in CPGE. (n) & (o) various (optical, neutron, electron tunneling) experiments with $k_x//M_z$ & $k_x \perp M_z$, respectively. (o) is not relevant to optical experiments. (p)-(q) antiferromagnetic states on a square lattice.

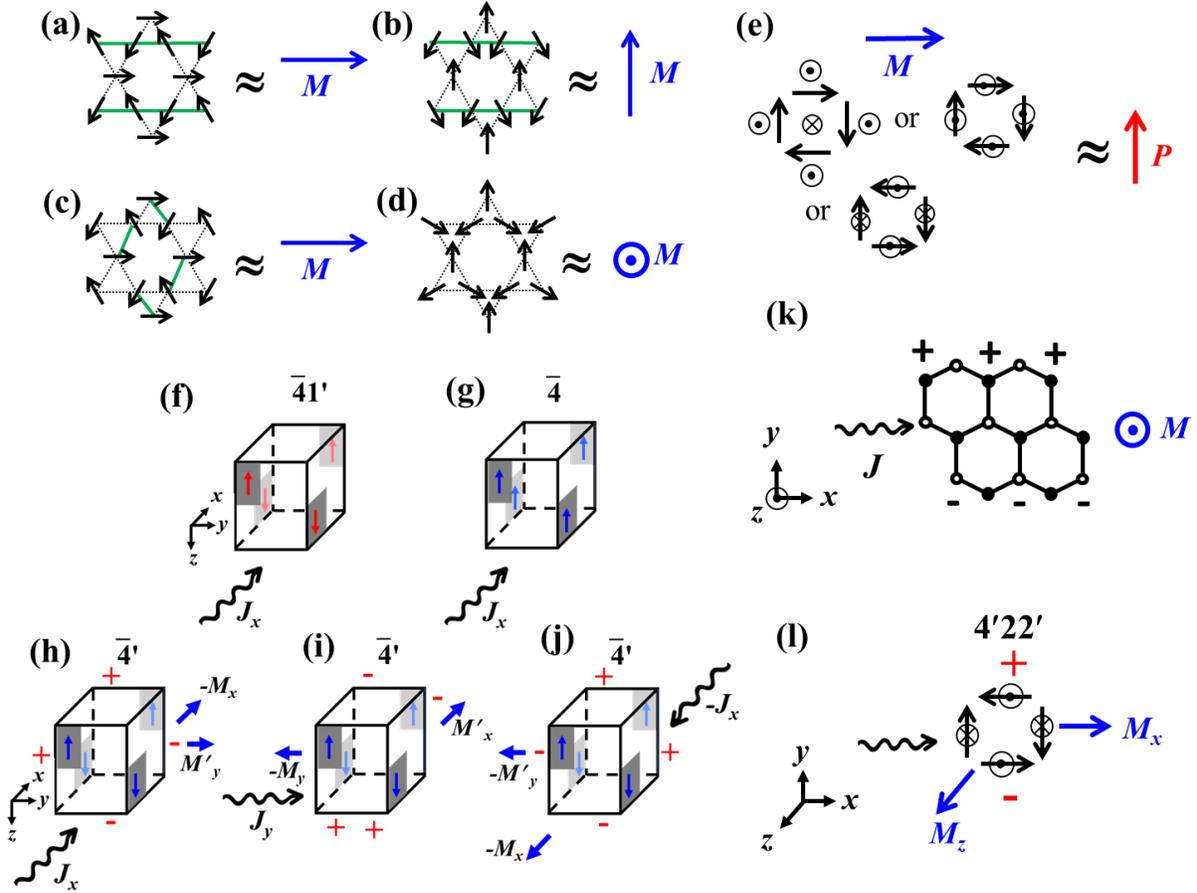

Fig. 3. Various specimens and experimental setups. (a)-(d) Four inversion-symmetric spin orders on kagome lattice. (a)-(b) are for Mn$_3$Sn (Magnetic Space Group (MSG): *Cmc'm'*; MPG: *mm'm'*) and Mn$_3$(Ge,Ga) (MSG: *Cm'cm'*; MPG: *m'mm'*) with orthorhombic cells. (c) is a kagome lattice with a monoclinic cell (MSG: *P2$_1$/n*; MPG: 2/*m*). (d) depicts one of three kagome lattices with ABC-type stacking in Mn$_3$(Rh,Ir,Pt) (MSG: *R$\bar{3}$m'*; MPG: $\bar{3}m'$). Green solid lines depict lattice distortions. (e) *P* represents the Hall effect of chiral spin textures (Left: Bloch-type skyrmion; Right: magnetic toroidal moment + canted moment; Bottom: magnetic quadrupole moment + alternating canted moment) due to magnetization induced by current. (f)-(j): Pictorial representation of three point groups exhibiting Odd-order AHE: (f) $\bar{4}$1' with unbroken ($\mathbf{R}_z,\bar{\mathbf{4}},\mathbf{T},\mathbf{R}_z\otimes\mathbf{T},\bar{\mathbf{4}}\otimes\mathbf{T}$): non-ferromagnetic & non-polar point group, (g) $\bar{4}$ with unbroken ($\mathbf{R}_z, \bar{\mathbf{4}}$): ferromagnetic & non-polar point groups, and (h)-(j) $\bar{4}'$ with unbroken ($\mathbf{R}_z, \bar{\mathbf{4}}'$) : non-ferromagnetic, non-polar, & non-chiral point group. In $\bar{4}'$, current along *x* can induce $M_x$ & $M_y$, Odd-order AHE voltage along *y*, and Even-order AHE voltage along *z*. (h) and (i) are linked through, e.g., $\mathbf{C}_{4z}\otimes\mathbf{I}$, and (h) and (j) are linked through, e.g., $\mathbf{R}_z$. (k) Experimental setup on $\bar{6}m$21' (e.g., monolayer *h*-BN or H-MoS$_2$) to measure $M_z$ induced by current along *x* and Even-order AHE. (l) "magnetic quadrupole + alternating canted moments" has no net moment, and corresponds with 4'22', and current along *x* on it can induce magnetizations along *x* and *z* and also Hall voltage along *y*.

**Table 1**. **Required broken symmetries and relevant MPGs for certain measurables.** When ferromagnetic point groups have net magnetic moment, then the net moment should be along $z$ for the AHE configuration in Fig. 2(a). All polar point groups do have polarization and the polarization should be along $y$ for the AHE configuration in Fig. 2(a) or off-diagonal current-induced **M** in Fig. 2(e). The subscriptions of $x$, $y$ and $z$ mean the relevant orientations: for example, Odd-order AHE$_{yx}$ means current along $x$ and Hall voltage along $y$, Even-order AHE$_{yx}$ means current along $x$ and Hall voltage along $y$, Diagonal current-induced $M_x$ is along $x$, and Optical activity$_z$ is along $z$. Note that Odd-order AHE$_{yx}$ = Odd-order AHE$_{xy}$ except sign in all ferromagnetic point groups. We emphasize that when we consider broken symmetries, we disregard or freely allow any rotations along the 1D direction for 1D phenomena such as Diagonal current-induced **M** and Optical activity (FR$_x$ means free rotation along $x$). For example, MPG $\bar{4}1'$ has broken **I**, but **I**⊗**C**$_{4z}$ is not broken, so it cannot exhibit Diagonal current-induced $M_z$; however, $\bar{4}1'$ can exhibit Diagonal current-induced $M_{x \text{ or } y}$. Odd-order AHE [1, 3, 5, 7, 9] and Even-order AHE [2, 4, 6, 8, 10] emerge as conjugate phenomena. For example, the set of broken {**I**⊗**T**,**T**,**M**$_x$,**M**$_y$,**R**$_x$,**R**$_y$,**C**$_{3x}$,**M**$_z$⊗**T**,**R**$_z$⊗**T**} for Odd-order AHE$_{yx}$ can be transform to broken {**I**⊗**T**,**I**,**R**$_x$⊗**T**,**R**$_y$⊗**T**,**R**$_x$,**R**$_y$,**C**$_{3x}$,**M**$_z$⊗**T**,**M**$_z$} through those operations (**T**→**I**, **M**$_x$→ **R**$_x$⊗**T**, **R**$_z$⊗**T**→**M**$_z$), corresponding to those for Even-order AHE$_{zx}$. Similarly, diagonal piezomagnetism [13, 15, 17] and diagonal piezoelectricity [14, 16, 18] are conjugate phenomena.

| Measurables | Required broken symmetries |
|---|---|
| [1] Odd-order AHE$_{yx}$ (48)<br>(Off-diagonal even-order current$_x$-induced $M_z$)<br>(Off-diagonal Piezomagnetism$_{zx}$) | {**I**⊗**T**,**T**,**M**$_x$,**M**$_y$,**R**$_x$,**R**$_y$,**C**$_{3x}$,**M**$_z$⊗**T**,**R**$_z$⊗**T**} |
| <ul><li>Ferromagnetic MPGs with $M_z$: 1, $\bar{1}$, 2$_{[z]}$, 2'$_{[x]}$, 2'$_{[y]}$, $m_{[z]}$, $m'_{[x]}$, $m'_{[y]}$, 2/$m_{[z]}$, 2'/$m'_{[x]}$, 2'/$m'_{[y]}$, 2'2'2, $m'm'2$, $m'm'm$, $m'2'm$, 2'$m'm$, 4, $\bar{4}$, 4/$m$, 42'2', 4$m'm'$, $\bar{4}$2$m'$, $\bar{4}m'$2', 4/$mm'm'$, 3, $\bar{3}$, 32'1, 312', 3$m'$1, 31$m'$, $\bar{3}m'$1, $\bar{3}$1$m'$, 6, $\bar{6}$, 6/$m$, 62'2', 6$m'm'$, $\bar{6}$2$m'$, $\bar{6}m'$2', 6/$mm'm'$</li><li>4', $\bar{4}'$, 4'/$m$, 4'2'2, 4'$m'm$, $\bar{4}'m'$2, $\bar{4}'$2'$m$, 4'/$mm'm$</li><li>23, $m\bar{3}$, 4'32', $\bar{4}'$3$m'$, $m\bar{3}m'$ for current along $xy$ or $yx$</li></ul> | |
| [2] Even-order AHE$_{zx}$ (48)<br>(Off-diagonal odd-order current$_x$-induced $M_y$)<br>(Off-diagonal Piezoelectrictiy$_{zx}$) | {**I**⊗**T**,**I**,**M**$_z$,**R**$_x$,**R**$_y$,**C**$_{3x}$,**M**$_z$⊗**T**,**R**$_x$⊗**T**,**R**$_y$⊗**T**} |
| <ul><li>Polar MPGs with $P_z$ : 1, 11', 2$_{[z]}$, 21'$_{[z]}$, 2'$_{[z]}$, $m_{[x]}$, $m_{[y]}$, $m'_{[x]}$, $m'_{[y]}$, $m1'_{[x]}$, $m1'_{[y]}$, $mm2$, $mm21'$, $m'm'2$, $m'm2'$, $mm'2'$, 4, 41', 4', 4$mm$, 4$mm1'$, 4'$mm'$, 4'$m'm$, 4$m'm'$, 3, 31', 3$m1$, 31$m$, 3$m11'$, 31$m1'$, 3$m'$1, 31$m'$, 6, 61', 6', 6$mm$, 6$mm1'$, 6'$mm'$, 6$m'm$, 6$m'm'$</li><li>$\bar{4}$, $\bar{4}'$, $\bar{4}1'$, $\bar{4}m2$, $\bar{4}m'2'$, $\bar{4}'m'2$, $\bar{4}'m2'$, $\bar{4}m21'$</li></ul> | |
| [3] Odd-order AHE$_{zx}$ (32)<br>(Off-diagonal even-order current$_x$-induced $M_y$)<br>(Off-diagonal Piezomagnetism$_{yx}$) | {**I**⊗**T**,**T**,**M**$_x$,**M**$_z$,**R**$_x$,**R**$_z$,**C**$_{3x}$,**M**$_y$⊗**T**,**R**$_y$⊗**T**} |
| <ul><li>Ferromagnetic MPGs with $M_y$: 1, $\bar{1}$, 2$_{[y]}$, 2'$_{[x]}$, 2'$_{[z]}$, $m_{[y]}$, $m'_{[x]}$, $m'_{[z]}$, 2/$m_{[y]}$, 2'/$m'_{[x]}$, 2'/$m'_{[z]}$, 2'22', $m'm2'$, 2'$mm'$, $m'2m'$, $m'mm'$</li><li>3, $\bar{3}$, 312, 32'1, 31$m$, 3$m'$1, $\bar{3}$1$m$, $\bar{3}m'$1, 6', $\bar{6}'$, 6'/$m'$, 6'$m'm$, 6'2'2, $\bar{6}'$2'$m$, $\bar{6}'m'$2, 6'/$m'm'm$</li></ul> | |
| [4] Even-order AHE$_{yx}$ (32) <br> (Off-diagonal odd-order current$_x$-induced $M_z$) <br> (Off-diagonal Piezoelectrictiy$_{yx}$) | {I⊗T,I,M$_y$,R$_x$,R$_z$,C$_{3x}$,M$_y$⊗T,R$_x$⊗T,R$_z$⊗T} |
| • Polar MPGs with $P_y$: 1, 11', 2$_{[y]}$, 21'$_{[y]}$, 2'$_{[y]}$, $m_{[x]}$, $m_{[z]}$, $m1'_{[x]}$, $m1'_{[z]}$, $m'_{[x]}$, $m'_{[z]}$, $m2m$, $m2m1'$, $m'2'm$, $m2'm'$, $m'2m'$ <br> • 3, 31', 312, 3121', 312', 3$m$1, 3$m$'1, 3$m$11', $\bar{6}$, $\bar{6}1'$, $\bar{6}'$, $\bar{6}m2$, $\bar{6}m21'$, $\bar{6}'m'2$, $\bar{6}'m2'$, $\bar{6}m'2'$ | |
| [5] Odd-order AHE$_{xy}$ (48) <br> (Off-diagonal even-order current$_y$-induced $M_z$) <br> (Off-diagonal Piezomagnetism$_{zy}$) | {I⊗T,T,M$_x$,M$_y$,R$_x$,R$_y$,C$_{3y}$,M$_z$⊗T,R$_z$⊗T} |
| • Ferromagnetic MPGs with $M_z$: 1, $\bar{1}$, 2$_{[z]}$, 2'$_{[x]}$, 2'$_{[y]}$, $m_{[z]}$, $m'_{[x]}$, $m'_{[y]}$, 2/$m_{[z]}$, 2'/$m'_{[x]}$, 2'/$m'_{[y]}$, 2'2'2, $m'm'2$, $m'm'm$, $m'2'm$, 2'$m'm$, 4, $\bar{4}$, 4/$m$, 42'2', 4$m'm'$, $\bar{4}2'm'$, $\bar{4}m'2'$, 4/$mm'm'$, 3, $\bar{3}$, 32'1, 312', 3$m$'1, 31$m$', $\bar{3}m'$1, $\bar{3}$1$m$', 6, $\bar{6}$, 6/$m$, 62'2', 6$m'm'$, $\bar{6}2'm'$, $\bar{6}m'2'$, 6/$mm'm'$ <br> • 4', $\bar{4}$', 4'/$m$, 4'2'2, 4'$m'm$, $\bar{4}'m'2$, $\bar{4}'2'm$, 4'/$mm'm$ <br> • 23, $m\bar{3}$, 4'32', $\bar{4}$'3$m$', $m\bar{3}m$' for current along $xy$ or $yx$ | |
| [6] Even-order AHE$_{zy}$ (48) <br> (Off-diagonal odd-order current$_y$-induced $M_x$) <br> (Off-diagonal Piezoelectrictiy$_{zy}$) | {I⊗T,I,M$_z$,R$_x$,R$_y$,C$_{3y}$,M$_z$⊗T,R$_x$⊗T,R$_y$⊗T} |
| • Polar MPGs with $P_z$ : 1, 11', 2$_{[z]}$, 21'$_{[z]}$, 2'$_{[z]}$, $m_{[x]}$, $m_{[y]}$, $m'_{[x]}$, $m'_{[y]}$, $m1'_{[x]}$, $m1'_{[y]}$, $mm2$, $mm21'$, $m'm'2$, $m'm2'$, $mm'2'$, 4, 41', 4', 4$mm$, 4$mm1'$, 4'$mm'$, 4'$m'm$, 4$m'm'$, 3, 31', 3$m$1, 31$m$, 3$m$11', 31$m$1', 3$m$'1, 31$m$', 6, 61', 6', 6$mm$, 6$mm1'$, 6'$mm'$, 6'$m'm$, 6$m'm'$ <br> • $\bar{4}$, $\bar{4}'$, $\bar{4}1'$, $\bar{4}m2$, $\bar{4}'m2'$, $\bar{4}'m'2$, $\bar{4}m'2'$, $\bar{4}m21'$ | |
| [7] Odd-order AHE$_{zy}$ (32) <br> (Off-diagonal even-order current$_y$-induced $M_x$) <br> (Off-diagonal Piezomagnetism$_{xy}$) | {I⊗T,T,M$_y$,M$_z$,R$_y$,R$_z$,C$_{3y}$,M$_x$⊗T,R$_x$⊗T} |
| • Ferromagnetic MPGs with $M_x$: 1, $\bar{1}$, 2$_{[x]}$, 2'$_{[y]}$, 2'$_{[z]}$, $m_{[x]}$, $m'_{[y]}$, $m'_{[z]}$, 2/$m_{[x]}$, 2'/$m'_{[y]}$, 2'/$m'_{[z]}$, 22'2', $mm'2'$, $m2'm'$, 2$m'm'$, $mm'm'$ <br> • 3, $\bar{3}$, 321, 312', 3$m$1, 31$m$', $\bar{3}m$1, $\bar{3}$1$m$', 6', $\bar{6}$', 6'/$m$', 6'22', 6'$mm$', $\bar{6}'m2$', $\bar{6}'2'm$, 6'/$m'mm$' | |
| [8] Even-order AHE$_{xy}$ (32) <br> (Off-diagonal odd-order current$_y$-induced $M_z$) <br> (Off-diagonal Piezoelectrictiy$_{xy}$) | {I⊗T,I,M$_x$,R$_y$,R$_z$,C$_{3y}$,M$_x$⊗T,R$_y$⊗T,R$_z$⊗T} |
| • Polar MPGs with $P_x$: 1, 11', 2$_{[x]}$, 21'$_{[x]}$, 2'$_{[x]}$, $m_{[y]}$, $m_{[z]}$, $m1'_{[y]}$, $m1'_{[z]}$, $m'_{[y]}$, $m'_{[z]}$, 2$mm$, 2$mm1'$, 2'$mm'$, 2'$m'm$, 2$m'm'$ <br> • 3, 31', 321, 3211', 32'1, 31$m$, 31$m$1', 31$m$', $\bar{6}$, $\bar{6}1'$, $\bar{6}'$, $\bar{6}2m$, $\bar{6}2m'$, $\bar{6}'2'm$, $\bar{6}2m1'$, $\bar{6}'2m'$ | |
| [9] Odd-order AHE$_{xz}$ (16) <br> (Off-diagonal even-order current$_z$-induced $M_y$) <br> (Off-diagonal Piezomagnetism$_{yz}$) | {I⊗T,T,M$_x$,M$_z$,R$_x$,R$_z$,C$_{3z}$,M$_y$⊗T,R$_y$⊗T} |
| • Ferromagnetic MPGs with $M_y$: 1, $\bar{1}$, 2$_{[y]}$, 2'$_{[x]}$, 2'$_{[z]}$, $m_{[y]}$, $m'_{[x]}$, $m'_{[z]}$, 2/$m_{[y]}$, 2'/$m'_{[x]}$, 2'/$m'_{[z]}$, 2'22', $m'm2$, 2'$mm$', $m'2m'$, $m'mm'$ | |

| [10] Even-order AHE$_{yz}$ (16) <br> (Off-diagonal odd-order current$_z$-induced $M_x$) <br> (Off-diagonal Piezoelectrictiy$_{yz}$) | {I⊗T,I,M$_y$,R$_x$,R$_z$,C$_{3z}$,M$_y$⊗T,R$_x$⊗T,R$_z$⊗T} |
|---|---|
| • Polar MPGs with $P_y$: 1, 11', 2$_{[y]}$, 21'$_{[y]}$, 2'$_{[y]}$, $m_{[x]}$, $m_{[z]}$, $m1'_{[x]}$, $m1'_{[z]}$, $m'_{[x]}$, $m'_{[z]}$, $m2m$, $m2m1'$, $m'2'm$, $m2m'$, $m'2m'$ | |
| [11] Odd-order AHE$_{yz}$ (16) <br> (Off-diagonal even-order current$_z$-induced $M_x$) <br> (Off-diagonal Piezomagnetism$_{xz}$) | {I⊗T,T,M$_y$,M$_z$,R$_y$,R$_z$,C$_{3z}$,M$_x$⊗T,R$_x$⊗T} |
| • Ferromagnetic MPGs with $M_x$: 1, $\bar{1}$, 2$_{[x]}$, 2'$_{[y]}$, 2'$_{[z]}$, $m_{[x]}$, $m'_{[y]}$, $m'_{[z]}$, 2/$m_{[x]}$, 2'/$m'_{[y]}$, 2'/$m'_{[z]}$, 22'2', $mm'2'$, $m2'm'$, $2m'm'$, $mm'm'$ | |
| [12] Even-order AHE$_{xz}$ (16) <br> (Off-diagonal odd-order current$_z$-induced $M_y$) <br> (Off-diagonal Piezoelectrictiy$_{xz}$) | {I⊗T,I,M$_x$,R$_y$,R$_z$,C$_{3z}$,M$_x$⊗T,R$_y$⊗T,R$_z$⊗T} |
| • Polar MPGs with $P_x$: 1, 11', 2$_{[x]}$, 21'$_{[x]}$, 2'$_{[x]}$, $m_{[y]}$, $m_{[z]}$, $m1'_{[y]}$, $m1'_{[z]}$, $m'_{[y]}$, $m'_{[z]}$, $2mm$, $2mm1'$, $2'mm'$, $2'm'm$, $2m'm'$ | |
| [13] Diagonal piezomagnetism$_x$ (FR$_x$) (32) <br> (Diagonal even-order current$_x$-induced $M_x$) (FR$_x$) | {I⊗T,T,M$_y$,M$_z$,R$_y$,R$_z$} |
| • Ferromagnetic MPGs with $M_x$: 1, $\bar{1}$, 2$_{[x]}$, 2'$_{[y]}$, 2'$_{[z]}$, $m_{[x]}$, $m'_{[y]}$, $m'_{[z]}$, 2/$m_{[x]}$, 2'/$m'_{[y]}$, 2'/$m'_{[z]}$, 22'2', $mm'2'$, $m2'm'$, $2m'm'$, $mm'm'$ <br> • 3, $\bar{3}$, 321, 312', $3m1$, $31m'$, $\bar{3}m1$, $\bar{3}1m'$, 6', $\bar{6}'$, 6'/$m'$, 6'22', 6'$mm'$, $\bar{6}'m2'$, $\bar{6}'2m'$, 6'/$m'mm'$ | |
| [14] Diagonal piezoelectricity$_x$ (FR$_x$) (32) | {I⊗T,I,R$_y$,R$_z$,R$_y$⊗T,R$_z$⊗T} |
| • Polar MPGs with $P_x$: 1, 11', 2$_{[x]}$, 21'$_{[x]}$, 2'$_{[x]}$, $m_{[y]}$, $m_{[z]}$, $m1'_{[y]}$, $m1'_{[z]}$, $m'_{[y]}$, $m'_{[z]}$, $2mm$, $2mm1'$, $2'mm'$, $2'm'm$, $2m'm'$ <br> • 3, 31', 321, 3211', 32'1, $31m$, $31m1'$, $31m'$, $\bar{6}$, $\bar{6}1'$, $\bar{6}'$, $\bar{6}2m$, $\bar{6}2'm'$, $\bar{6}'2'm$, $\bar{6}2m1'$, $\bar{6}'2m'$ | |
| [15] Diagonal piezomagnetism$_y$ (FR$_y$) (32) <br> (Diagonal even-order current$_y$-induced $M_y$) (FR$_y$) | {I⊗T,T,M$_x$,M$_z$,R$_x$,R$_z$} |
| • Ferromagnetic MPGs with $M_y$: 1, $\bar{1}$, 2$_{[y]}$, 2'$_{[x]}$, 2'$_{[z]}$, $m_{[y]}$, $m'_{[x]}$, $m'_{[z]}$, 2/$m_{[y]}$, 2'/$m'_{[x]}$, 2'/$m'_{[z]}$, 2'22', $m'm2'$, $2'mm'$, $m'2m'$, $m'mm'$ <br> • 3, $\bar{3}$, 312, 32'1, $31m$, $3m'1$, $\bar{3}1m$, $\bar{3}m'1$, 6', $\bar{6}'$, 6'/$m'$, 6'$m'm$, 6'2'2, $\bar{6}'2'm$, $\bar{6}'m'2$, 6'/$m'm'm$ | |
| [16] Diagonal piezoelectricity$_y$ (FR$_y$) (32) | {I⊗T,I,R$_x$,R$_z$,R$_x$⊗T,R$_z$⊗T} |
| • Polar MPGs with $P_y$: 1, 11', 2$_{[y]}$, 21'$_{[y]}$, 2'$_{[y]}$, $m_{[x]}$, $m_{[z]}$, $m1'_{[x]}$, $m1'_{[z]}$, $m'_{[x]}$, $m'_{[z]}$, $m2m$, $m2m1'$, $m'2'm$, $m2m'$, $m'2m'$ <br> • 3, 31', 312, 3121', 312', $3m1$, $3m'1$, $3m11'$, $\bar{6}$, $\bar{6}1'$, $\bar{6}'$, $\bar{6}m2$, $\bar{6}m21'$, $\bar{6}'m'2$, $\bar{6}'m2'$, $\bar{6}m2'$ | |
| [17] Diagonal piezomagnetism$_z$ (FR$_z$) (40) <br> (Diagonal even-order current$_z$-induced $M_z$) (FR$_z$) | {I⊗T,T,M$_x$,M$_y$,R$_x$,R$_y$} |
| • Ferromagnetic MPGs with $M_z$: 1, $\bar{1}$, 2$_{[z]}$, 2'$_{[x]}$, 2'$_{[y]}$, $m_{[z]}$, $m'_{[x]}$, $m'_{[y]}$, 2/$m_{[z]}$, 2'/$m'_{[x]}$, 2'/$m'_{[y]}$, 2'2'2, $m'm'2$, $m'm'm$, $m'2'm$, $2'm'm$, 4, $\bar{4}$, 4/$m$, 42'2', $4m'm'$, $\bar{4}2'm'$, $\bar{4}m'2'$, 4/$mm'm'$, 3, $\bar{3}$, 32'1, 312', $31m'$, $3m'1$, $\bar{3}m'1$, $\bar{3}1m'$, 6, $\bar{6}$, 6/$m$, 62'2', $6m'm'$, $\bar{6}2'm'$, $\bar{6}m'2'$, 6/$mm'm'$ | |
| [18] Diagonal piezoelectricity$_z$ (FR$_z$) (40) | **{I⊗T,I,R$_x$,R$_y$,R$_x$⊗T,R$_y$⊗T}** |

- Polar MPGs with $P_z$ : 1, 11', 2$_{[z]}$, 21'$_{[z]}$, 2'$_{[z]}$, $m_{[x]}$, $m_{[y]}$, $m'_{[x]}$, $m'_{[y]}$, $m1'_{[x]}$, $m1'_{[y]}$, $mm2$, $mm21'$, $m'm'2$, $m'm2'$, $mm'2'$, 4, 41', 4', $4mm$, $4mm1'$, $4'mm'$, $4'm'm$, $4m'm'$, 3, 31', $3m1$, $31m$, $3m11'$, $31m1'$, $3m'1$, $31m'$, 6, 61', 6', $6mm$, $6mm1'$, $6'mm'$, $6'm'm$, $6m'm'$

| | |
|---|---|
| [19] Directional nonreciprocity$_x$ in $H_z$ (32) nonreciprocity$_z$ in $H_x$ | **{I⊗T,I,M$_y$,R$_x$,R$_z$,M$_y$⊗T,R$_x$⊗T,R$_z$⊗T}** |

- Polar MPGs with $P_y$: 1, 11', 2$_{[y]}$, 21'$_{[y]}$, 2'$_{[y]}$, $m_{[x]}$, $m_{[z]}$, $m1'_{[x]}$, $m1'_{[z]}$, $m'_{[x]}$, $m'_{[z]}$, $m2m$, $m2m1'$, $m'2m$, $m2'm'$, $m'2m'$
- 3, 31', 312, 3121', 312', $3m1$, $3m'1$, $3m11'$, $\bar{6}$, $\bar{6}1'$, $\bar{6}'$, $\bar{6}m2$, $\bar{6}m21'$, $\bar{6}'m'2$, $\bar{6}m'2'$, $\bar{6}'m2'$

| | |
|---|---|
| [20] Directional nonreciprocity$_x$ in $H_y$ (48) nonreciprocity$_y$ in $H_x$ | **{I⊗T,I,M$_z$,R$_x$,R$_y$,M$_z$⊗T,R$_x$⊗T,R$_y$⊗T}** |

- Polar MPGs with $P_z$ : 1, 11', 2$_{[z]}$, 21'$_{[z]}$, 2'$_{[z]}$, $m_{[x]}$, $m_{[y]}$, $m'_{[x]}$, $m'_{[y]}$, $m1'_{[x]}$, $m1'_{[y]}$, $mm2$, $mm21'$, $m'm'2$, $m'm2'$, $mm'2'$, 4, 41', 4', $4mm$, $4mm1'$, $4'mm'$, $4'm'm$, $4m'm'$, 3, 31', $3m1$, $31m$, $3m11'$, $31m1'$, $3m'1$, $31m'$, 6, 61', 6', $6mm$, $6mm1'$, $6'mm'$, $6'm'm$, $6m'm'$
- $\bar{4}$, $\bar{4}'$, $\bar{4}1'$, $\bar{4}m2$, $\bar{4}m'2'$, $\bar{4}'m'2$, $\bar{4}'m2'$, $\bar{4}m21'$

| | |
|---|---|
| [21] Directional nonreciprocity$_y$ in $H_z$ (32) nonreciprocity$_z$ in $H_y$ | **{I⊗T,I,M$_x$,R$_y$,R$_z$,M$_x$⊗T,R$_y$⊗T,R$_z$⊗T}** |

- Polar MPGs with $P_x$: 1, 11', 2$_{[x]}$, 21'$_{[x]}$, 2'$_{[x]}$, $m_{[y]}$, $m_{[z]}$, $m1'_{[y]}$, $m1'_{[z]}$, $m'_{[y]}$, m'$_{[z]}$, $2mm$, $2mm1'$, $2'mm'$, $2'm'm$, $2m'm'$
- 3, 31', 321, 3211', 32'1, $31m$, $31m1'$, $31m'$, $\bar{6}$, $\bar{6}1'$, $\bar{6}'$, $\bar{6}2m$, $\bar{6}2'm'$, $\bar{6}'2'm$, $\bar{6}2m1'$, $\bar{6}'2m'$

| | |
|---|---|
| [22] Diagonal odd-order current$_x$-induced $M_x$ (FR$_x$) (53) (Directional nonreciprocity$_x$ in $H_x$) | **{I⊗T,I,M$_y$,M$_z$,M$_y$⊗T,M$_z$⊗T}** |

- Chiral MPGs: 1, 11', 2$_{[x]}$, 2$_{[y]}$, 2$_{[z]}$, 21'$_{[x]}$, 21'$_{[y]}$, 21'$_{[z]}$, 2'$_{[x]}$, 2'$_{[y]}$, 2'$_{[z]}$, 222, 2221', 2'2'2, 22'2', 2'22', 4, 41', 4', 422, 4221', 4'2'2, 4'22', 42'2', 3, 31', 321, 312, 3211', 3121', 32'1, 312', 6, 61', 6', 622, 6221', 6'22', 6'2'2, 62'2', 23, 231', 432, 4321', 4'32'
- $\bar{4}$, $\bar{4}1'$, $\bar{4}'$, $\bar{4}2m$, $\bar{4}2m1'$, $\bar{4}'2'm$, $\bar{4}2'm'$, $\bar{4}'2m'$

| | |
|---|---|
| [23] Diagonal odd-order current$_y$-induced $M_y$ (FR$_y$) (53) (Directional nonreciprocity$_y$ in $H_y$) | **{I⊗T,I,M$_x$,M$_z$,M$_x$⊗T,M$_z$⊗T}** |

- Chiral MPGs: 1, 11', 2$_{[x]}$, 2$_{[y]}$, 2$_{[z]}$, 21'$_{[x]}$, 21'$_{[y]}$, 21'$_{[z]}$, 2'$_{[x]}$, 2'$_{[y]}$, 2'$_{[z]}$, 222, 2221', 2'2'2, 22'2', 2'22', 4, 41', 4', 422, 4221', 4'2'2, 4'22', 42'2', 3, 31', 321, 312, 3211', 3121', 32'1, 312', 6, 61', 6', 622, 6221', 6'22', 6'2'2, 62'2', 23, 231', 432, 4321', 4'32'
- $\bar{4}$, $\bar{4}1'$, $\bar{4}'$, $\bar{4}2m$, $\bar{4}2m1'$, $\bar{4}'2'm$, $\bar{4}2'm'$, $\bar{4}'2m'$

| | |
|---|---|
| [24] Diagonal odd-order current$_z$-induced $M_z$ (FR$_z$) (45) (Directional nonreciprocity$_z$ in $H_z$) | **{I⊗T,I,M$_x$,M$_y$,M$_x$⊗T,M$_y$⊗T}** |

- Chiral MPGs: 1, 11', 2$_{[x]}$, 2$_{[y]}$, 2$_{[z]}$, 21'$_{[x]}$, 21'$_{[y]}$, 21'$_{[z]}$, 2'$_{[x]}$, 2'$_{[y]}$, 2'$_{[z]}$, 222, 2221', 2'2'2, 22'2', 2'22', 4, 41', 4', 422, 4221', 4'2'2, 4'22', 42'2', 3, 31', 321, 312, 3211', 3121', 32'1, 312', 6, 61', 6', 622, 6221', 6'22', 6'2'2, 62'2', 23, 231', 432, 4321', 4'32'

| [25] Optical activity$_x$ (75) (Faraday effect, CPGE, circular dichroism, Natural optical activity) with $k_x//M_x$ (FR$_x$) | {I⊗T,M$_y$,M$_z$} |
|---|---|

- Ferromagnetic MPGs with $M_x$: $\bar{1}$, $m_{[x]}$, $m'_{[y]}$, $m'_{[z]}$, $2/m_{[x]}$, $2'/m'_{[y]}$, $2'/m'_{[z]}$, $mm'2'$, $m2'm'$, $2m'm'$, $mm'm'$
- Chiral MPGs: 11', $2_{[y]}$, $2_{[z]}$, $21'_{[x]}$, $21'_{[y]}$, $21'_{[z]}$, $2'_{[x]}$, 222, 2221', 2'2'2, 2'22', 4, 41', 4', 422, 4221', 4'2'2, 4'22', 42'2', 3, 31', 321, 312, 3211', 3121', 32'1, 312', 6, 61', 6', 622, 6221', 6'22', 6'2'2, 62'2', 23, 231', 432, 4321', 4'32'
- $M_x$ &Chiral: 1, $2_{[x]}$, $2'_{[y]}$, $2'_{[z]}$, 22'2'
- $\bar{3}$, $3m1$, $31m'$, $\bar{3}m1$, $\bar{3}1m'$, $\bar{6}$, $6'/m'$, $6'mm'$, $\bar{6}'m2'$, $\bar{6}'2m'$, $6'/m'mm'$
- $\bar{4}$, $\bar{4}1'$, $\bar{4}'$, $\bar{4}2m$, $\bar{4}2m1'$, $\bar{4}'2m$, $\bar{4}'2m'$, $\bar{4}2'm'$

| [26] Optical activity$_y$ (75) (Faraday effect, CPGE, circular dichroism, Natural optical activity) with $k_y//M_y$ (FR$_y$) | {I⊗T,M$_x$,M$_z$} |
|---|---|

- Ferromagnetic MPGs with $M_y$: $\bar{1}$, $m_{[y]}$, $m'_{[x]}$, $m'_{[z]}$, $2/m_{[y]}$, $2'/m'_{[x]}$, $2'/m'_{[z]}$, $m'm2'$, $2'mm'$, $m'2m'$, $m'mm'$
- Chiral MPGs: 11', $2_{[x]}$, $2_{[z]}$, $21'_{[x]}$, $21'_{[y]}$, $21'_{[z]}$, $2'_{[y]}$, 222, 2221', 2'2'2, 22'2', 4, 41', 4', 422, 4221', 4'2'2, 4'22', 42'2', 3, 31', 321, 312, 3211', 3121', 32'1, 312', 6, 61', 6', 622, 6221', 6'22', 6'2'2, 62'2', 23, 231', 432, 4321', 4'32'
- $M_y$ &Chiral: 1, $2_{[y]}$, $2'_{[x]}$, $2'_{[z]}$, 2'22'
- $\bar{3}$, $31m$, $3m'1$, $\bar{3}1m$, $\bar{3}m'1$, $\bar{6}$, $6'/m'$, $6'm'm$, $\bar{6}'2m'$, $\bar{6}'m'2$, $6'/m'm'm$
- $\bar{4}$, $\bar{4}1'$, $\bar{4}'$, $\bar{4}2m$, $\bar{4}2m1'$, $\bar{4}'2m$, $\bar{4}'2m'$, $\bar{4}2'm'$

| [27] Optical activity$_z$ (73) (Faraday effect, CPGE, circular dichroism, Natural optical activity) with $k_z//M_z$ (FR$_z$) | {I⊗T,M$_x$,M$_y$} |
|---|---|

- Ferromagnetic MPGs with $M_z$: $\bar{1}$, $m_{[z]}$, $m'_{[x]}$, $m'_{[y]}$, $2/m_{[z]}$, $2'/m'_{[x]}$, $2'/m'_{[y]}$, $m'm'2$, $m'm'm$, $m'2'm$, $2'm'm$, $\bar{4}$, $4/m$, $4m'm'$, $\bar{4}2'm'$, $\bar{4}m'2'$, $4/mm'm'$, $\bar{3}$, $3m'1$, $31m'$, $\bar{3}m'1$, $\bar{3}1m'$, $\bar{6}$, $6/m$, $6m'm'$, $\bar{6}2'm'$, $\bar{6}m'2'$, $6/mm'm'$
- Chiral MPGs: 11', $2_{[x]}$, $2_{[y]}$, $21'_{[x]}$, $21'_{[y]}$, $21'_{[z]}$, $2'_{[z]}$, 222, 2221', 22'2', 2'22', 41', 4', 422, 4221', 4'2'2, 4'22', 31', 321, 312, 3211', 3121', 61', 6', 622, 6221', 6'22', 6'2'2, 23, 231', 432, 4321', 4'32'
- $M_z$ &Chiral: 1, $2_{[z]}$, $2'_{[x]}$, $2'_{[y]}$, 2'2'2, 4, 42'2', 3, 32'1, 312', 6, 62'2'

| [28] Off-diagonal activity with $k_x \perp M_z$ (70) | {I⊗T,M$_y$,R$_x$,R$_z$⊗T,C$_{3x}$} |
|---|---|

- Ferromagnetic MPGs with $M_z$: $\bar{1}$, $2_{[z]}$, $2'_{[x]}$, $m'_{[y]}$, $2/m_{[z]}$, $2'/m'_{[x]}$, $2'/m'_{[y]}$, 2'2'2, $m'm'2$, $m'm'm$, $2'm'm$, 4, $\bar{4}$, $4/m$, 42'2', $4m'm'$, $\bar{4}2'm'$, $\bar{4}m'2'$, $4/mm'm'$, 3, $\bar{3}$, 32'1, 312', $3m'1$, $31m'$, $\bar{3}m'1$, $\bar{3}1m'$, 6, $\bar{6}$, $6/m$, 62'2', $6m'm'$, $\bar{6}2'm'$, $\bar{6}m'2'$, $6/mm'm'$
- Polar MPGs with $P_y$: 11', $2_{[y]}$, $21'_{[y]}$, $m_{[x]}$, $m1'_{[x]}$, $m1'_{[z]}$, $m'_{[z]}$, $m2m$, $m2m1'$, $m'2'm$, $m2'm'$
- $M_z$ &$P_y$: 1, $2'_{[y]}$, $m_{[z]}$, $m'_{[x]}$, $m'2m$
- 4', $\bar{4}'$, 4'/m, 4'2'2, 4'm'm, $\bar{4}'m2$, $\bar{4}'2'm$, 4'/mm'm
- 31', 312, 3121', $3m1$, $3m11'$, $\bar{6}1'$, $\bar{6}$, $\bar{6}m2$, $\bar{6}m21'$, $\bar{6}'m'2$, $\bar{6}'m2'$

| [29] Off-diagonal activity with $k_x \perp M_y$ (70) | $\{I \otimes T, M_z, R_x, R_y \otimes T, C_{3x}\}$ |
|---|---|

- Ferromagnetic MPGs with $M_y$: $\bar{1}$, $2_{[y]}$, $2'_{[x]}$, $m'_{[z]}$, $2/m_{[y]}$, $2'/m'_{[x]}$, $2'/m'_{[z]}$, $2'22'$, $2'mm'$, $m'2m'$, $m'mm'$
- Polar MPGs with $P_z$: $11'$, $2_{[z]}$, $21'_{[z]}$, $m_{[x]}$, $m'_{[y]}$, $m1'_{[x]}$, $m1'_{[y]}$, $mm2$, $mm21'$, $m'm'2$, $mm'2'$, $4$, $41'$, $4'$, $4mm$, $4mm1'$, $4'mm$, $4'm'm$, $4m'm'$, $3$, $31'$, $3m1$, $31m$, $3m11'$, $31m1'$, $3m'1$, $31m'$, $6$, $61'$, $6'$, $6mm$, $6mm1'$, $6'mm'$, $6'm'm$, $6m'm'$
- $M_y$ & $P_z$: $1$, $2'_{[z]}$, $m_{[y]}$, $m'_{[x]}$, $m'm2'$
- $\bar{3}$, $312$, $32'1$, $\bar{3}1m$, $\bar{3}m'1$, $\bar{6}$, $6'/m'$, $6'2'2$, $\bar{6}'2'm$, $\bar{6}'m'2$, $6'/m'm'm$
- $\bar{4}$, $\bar{4}'$, $\bar{4}1'$, $\bar{4}m2$, $\bar{4}m'2'$, $\bar{4}'m'2$, $\bar{4}'m2'$, $\bar{4}m21'$

| [30] Off-diagonal activity with $k_y \perp M_z$ (70) | $\{I \otimes T, M_x, R_y, R_z \otimes T, C_{3y}\}$ |
|---|---|

- Ferromagnetic MPGs with $M_z$: $\bar{1}$, $2_{[z]}$, $2'_{[y]}$, $m'_{[x]}$, $2/m_{[z]}$, $2'/m'_{[x]}$, $2'/m'_{[y]}$, $2'2'2$, $m'm'2$, $m'm'm$, $m'2m'$, $4$, $\bar{4}$, $4/m$, $42'2'$, $4m'm'$, $\bar{4}2'm'$, $\bar{4}m'2'$, $4/mm'm'$, $3$, $\bar{3}$, $32'1$, $312$, $3m'1$, $31m'$, $\bar{3}m'1$, $\bar{3}1m'$, $6$, $\bar{6}$, $6/m$, $62'2'$, $6m'm'$, $\bar{6}2'm'$, $\bar{6}m'2'$, $6/mm'm'$
- Polar MPGs with $P_x$: $11'$, $2_{[x]}$, $21'_{[x]}$, $m_{[y]}$, $m1'_{[y]}$, $m1'_{[z]}$, $m'_{[z]}$, $2mm$, $2mm1'$, $2'mm'$, $2m'm'$
- $M_z$ & $P_x$: $1$, $2'_{[x]}$, $m_{[z]}$, $m'_{[y]}$, $2'm'm$
- $4'$, $\bar{4}'$, $4'/m$, $4'2'2$, $4'm'm$, $\bar{4}'m2$, $\bar{4}'2'm$, $4'/mm'm$
- $31'$, $321$, $3211'$, $31m$, $31m1'$, $\bar{6}1'$, $\bar{6}'$, $\bar{6}62m$, $\bar{6}'2'm$, $\bar{6}2m1'$, $\bar{6}'2m'$

| [31] Off-diagonal activity with $k_y \perp M_x$ (70) | $\{I \otimes T, M_z, R_y, R_x \otimes T, C_{3y}\}$ |
|---|---|

- Ferromagnetic MPGs with $M_x$: $\bar{1}$, $2_{[x]}$, $2'_{[y]}$, $m'_{[z]}$, $2/m_{[x]}$, $2'/m'_{[y]}$, $2'/m'_{[z]}$, $22'2'$, $m2'm'$, $2m'm'$, $mm'm'$
- Polar MPGs with $P_z$: $11'$, $2_{[z]}$, $21'_{[z]}$, $m_{[y]}$, $m'_{[x]}$, $m1'_{[x]}$, $m1'_{[y]}$, $mm2$, $mm21'$, $m'm'2$, $m'm2'$, $4$, $41'$, $4'$, $4mm$, $4mm1'$, $4'mm$, $4'm'm$, $4m'm'$, $3$, $31'$, $3m1$, $31m$, $3m11'$, $31m1'$, $3m'1$, $31m'$, $6$, $61'$, $6'$, $6mm$, $6mm1'$, $6'mm'$, $6'm'm$, $6m'm'$
- $M_x$ & $P_z$: $1$, $2'_{[z]}$, $m_{[x]}$, $m'_{[y]}$, $mm'2'$
- $\bar{3}$, $321$, $312'$, $\bar{3}m1$, $\bar{3}1m'$, $\bar{6}'$, $6'/m'$, $6'22'$, $\bar{6}'m2'$, $\bar{6}'2m'$, $6'/m'mm$
- $\bar{4}$, $\bar{4}'$, $\bar{4}1'$, $\bar{4}m2$, $\bar{4}m'2'$, $\bar{4}'m'2$, $\bar{4}'m2'$, $\bar{4}m21'$

| [32] Off-diagonal activity with $k_z \perp M_y$ (27) | $\{I \otimes T, M_x, R_z, R_y \otimes T, C_{3z}\}$ |
|---|---|

- Ferromagnetic MPGs with $M_y$: $\bar{1}$, $2_{[y]}$, $2'_{[z]}$, $m'_{[x]}$, $2/m_{[y]}$, $2'/m'_{[x]}$, $2'/m'_{[z]}$, $2'22'$, $m'm2$, $m'2m'$, $m'mm'$
- Polar MPGs with $P_x$: $11'$, $2_{[x]}$, $21'_{[x]}$, $m_{[z]}$, $m1'_{[y]}$, $m1'_{[z]}$, $m'_{[y]}$, $2mm$, $2mm1'$, $2'mm'$, $2m'm'$
- $M_y$ & $P_x$: $1$, $2'_{[x]}$, $m_{[y]}$, $m'_{[z]}$, $2'mm$

| [33] Off-diagonal activity with $k_z \perp M_x$ (27) | $\{I \otimes T, M_y, R_z, R_x \otimes T, C_{3z}\}$ |
|---|---|

- Ferromagnetic MPGs with $M_x$: $\bar{1}$, $2_{[x]}$, $2'_{[z]}$, $m'_{[y]}$, $2/m_{[x]}$, $2'/m'_{[y]}$, $2'/m'_{[z]}$, $22'2'$, $mm'2$, $2m'm'$, $mm'm'$
- Polar MPGs with $P_y$: $11'$, $2_{[y]}$, $21'_{[y]}$, $m_{[z]}$, $m1'_{[x]}$, $m1'_{[z]}$, $m'_{[x]}$, $m2m$, $m2m1'$, $m'2'm$, $m2m'$,
- $M_x$ & $P_y$: $1$, $2'_{[y]}$, $m_{[x]}$, $m'_{[z]}$, $m'2m'$

| [34] Shear stress-induced piezoelectricity$_x$ (53) | {I⊗T,I,M$_x$,M$_y$,M$_z$,M$_x$⊗T,M$_y$⊗T,M$_z$⊗T, C$_{3x}$,C$_{3x}$⊗T,C$_{4x}$,C$_{4x}$⊗T } |
|---|---|
| <ul><li>Chiral MPGs: 1, 11', 2$_{[x]}$, 2$_{[y]}$, 2$_{[z]}$, 21'$_{[x]}$, 21'$_{[y]}$, 21'$_{[z]}$, 2'$_{[x]}$, 2'$_{[y]}$, 2'$_{[z]}$, 222, 2221', 2'2'2, 22'2', 2'22', 4, 41', 4', 422, 4221', 4'2'2, 4'22', 42'2', 3, 31', 321, 312, 3211', 3121', 32'1, 312', 6, 61', 6', 622, 6221', 6'22', 6'2'2, 62'2', 23, 231'</li><li>$\bar{4}$, $\bar{4}$1', $\bar{4}$', $\bar{4}$2m, $\bar{4}$2m1', $\bar{4}$'2'm, $\bar{4}$'2m', $\bar{4}$2'm'</li><li>$\bar{4}$3m, $\bar{4}$3m1', $\bar{4}$'3m'</li></ul> | |
| [35] Shear stress-induced piezoelectricity$_y$ (53) | {I⊗T,I,M$_x$,M$_y$,M$_z$,M$_x$⊗T,M$_y$⊗T,M$_z$⊗T, C$_{3y}$,C$_{3y}$⊗T,C$_{4y}$,C$_{4y}$⊗T } |
| <ul><li>Chiral MPGs: 1, 11', 2$_{[x]}$, 2$_{[y]}$, 2$_{[z]}$, 21'$_{[x]}$, 21'$_{[y]}$, 21'$_{[z]}$, 2'$_{[x]}$, 2'$_{[y]}$, 2'$_{[z]}$, 222, 2221', 2'2'2, 22'2', 2'22', 4, 41', 4', 422, 4221', 4'2'2, 4'22', 42'2', 3, 31', 321, 312, 3211', 3121', 32'1, 312', 6, 61', 6', 622, 6221', 6'22', 6'2'2, 62'2', 23, 231'</li><li>$\bar{4}$, $\bar{4}$1', $\bar{4}$', $\bar{4}$2m, $\bar{4}$2m1', $\bar{4}$'2'm, $\bar{4}$'2m', $\bar{4}$2'm'</li><li>$\bar{4}$3m, $\bar{4}$3m1', $\bar{4}$'3m'</li></ul> | |
| [36] Shear stress-induced piezoelectricity$_z$ (29) | {I⊗T,I,M$_x$,M$_y$,M$_z$,M$_x$⊗T,M$_y$⊗T,M$_z$⊗T, C$_{3z}$,C$_{3z}$⊗T,C$_{4z}$,C$_{4z}$⊗T} |
| <ul><li>Chiral MPGs: 1, 11', 2$_{[x]}$, 2$_{[y]}$, 2$_{[z]}$, 21'$_{[x]}$, 21'$_{[y]}$, 21'$_{[z]}$, 2'$_{[x]}$, 2'$_{[y]}$, 2'$_{[z]}$, 222, 2221', 2'2'2, 22'2', 2'22', 23, 231'</li><li>$\bar{4}$, $\bar{4}$1', $\bar{4}$', $\bar{4}$2m, $\bar{4}$2m1', $\bar{4}$'2'm, $\bar{4}$'2m', $\bar{4}$2'm'</li><li>$\bar{4}$3m, $\bar{4}$3m1', $\bar{4}$'3m'</li></ul> | |

**REFERENCES:**

[1] Ion I. Geru, Time-reversal symmetry (Springer Nature Switzerland, 2018)

[2] C. N. Yang, Law of parity conservation and other symmetry laws, Science **127**, 565 (1958).

[3] A. Robinson, 1980 Nobel Prize in Physics to Cronin and Fitch, Science **210**, 619 (1980).

[4] C. M. Bender and S. Boettcher, Real spectra in non-Hermitian hamiltonians having PT symmetry, Phys. Rev. Lett. **80**, 5243 (1998).

[5] C. M. Bender, Introduction to PT-Symmetric quantum theory, Contemp. Phys. **46**, 277 (2005).

[6] R. El-Ganainy, K. G. Makris, M. Khajavikhan, Z. H. Musslimani, S. Rotter and D. N. Christodoulides, Non-Hermitian physics and PT symmetry, Nat. Phys. **14**, 11 (2018).

[7] Y. Wu, W. Liu, J. Geng, X. Song, X. Ye, C.-K. Duan, X. Rong and J. Du, Observation of parity-time symmetry breaking in a single spin system, Science **364**, 878 (2019).

[8] Ş. K. Özdemir, S. Rotter, F. Nori and L. Yang, Parity–time symmetry and exceptional points in photonics, Nat. Mater. **18**, 783 (2019).

[9] A. Manchon, H. C. Koo, J. Nitta, S. M. Frolov and R. A. Duine, New perspectives for Rashba spin-orbit coupling, Nat. Mater. **14**, 871 (2015).

[10] A. Bansil, H. Lin and T. Das, Colloquium: Topological band theory, Revs. Mod. Phys. **88**, 021004 (2016).

[11] A. Soumyanarayanan, N. Reyren, A. Fert and C. Panagopoulos, Emergent phenomena induced by spin–orbit coupling at surfaces and interfaces, Nature **539**, 509 (2016).

[12] D. Hsieh, Y. Xia, L. Wray, D. Qian, A. Pal, J. H. Dil, J. Osterwalder, F. Meier, G. Bihlmayer, C. L. Kane, Y. S. Hor, R. J. Cava and M. Z. Hasan, Observation of unconventional quantum spin textures in topological insulators, Science **323**, 919 (2009).
[13] H. Zhu, J. Yi, M.-Y. Li, J. Xiao, L. Zhang, C.-W. Yang, R. A. Kaindl, L.-J. Li, Y. Wang and X. Zhang, Observation of chiral phonons, Science **359**, 579 (2018).
[14] D. M. Juraschek and N. A. Spaldin, Orbital magnetic moments of phonons, Phys. Rev. Mater. **3**, 064405 (2019).
[15] S. R. Tauchert, M. Volkov, D. Ehberger, D. Kazenwadel, M. Evers, H. Lange, A. Donges, A. Book, W. Kreuzpaintner, U. Nowak and P. Baum, Polarized phonons carry angular momentum in ultrafast demagnetization, Nature **602**, 73 (2022).
[16] P. Curie, Sur la symétrie dans les phénomènes physiques, symétrie d'un champ électrique et d'un champ magnétique, J. Phys. Theor. Appl **3**, 393 (1894).
[17] O. Meyer and F. Neumann, *Vorlesungen über die Theorie der Elasticität der Festen Körper und des Lichtäthers: Gehalten an der Universität Königsberg: von Dr. Franz Neumann. Hrsg. von Dr. Oskar Emil Meyer, mit Figuren Im Text*. (B. G. Teubner, Leipzig, 1885).
[18] S.-W. Cheong, S. Lim, K. Du and F.-T. Huang, Permutable SOS (symmetry operational similarity), NPJ Quantum Mater. **6**, 58 (2021).
[19] S.-W. Cheong, SOS: symmetry-operational similarity, NPJ Quantum Mater. **4**, 53 (2019).
[20] S.-W. Cheong, Trompe L'oeil Ferromagnetism, NPJ Quantum Mater. **5**, 37 (2020).
[21] S.-W. Cheong, F.-T. Huang and K. Minhyong, Linking emergent phenomena and broken symmetries through one-dimensional objects and their dot/cross products, Rep. Prog. Phys. **85**, 124501 (2022).
[22] W. Opechowski and R. Guccione, *Magnetism.* edited by G. T. Rado and H. Suhl (Academic Press, New York, 1965).
[23] N. Belov, N. Neronova and T. Smirnova, Soviet Phys. Cryst. **2**, 311 (1957).
[24] A. S. Borovik-Romanov and H. Grimmer, International Tables for Crystallography Vol. D, edited by A. Authier (International Union of Crystallography by Kluwer Academic Publishers, London, 2010) page 110-111.
[25] H. Schmid, Some symmetry aspects of ferroics and single phase multiferroics, J. Phys. Condens. Matter **20**, 434201 (2008).
[26] R. R. Birss, Symmetry and magnetism. (North-Holland Pub. Co., Amsterdam, 1964).
[27] L. A. Shuvalov, Modern crystallography. IV. Physical properties of crystals. (Springer England, 1989).
[28] T. Furukawa, Y. Watanabe, N. Ogasawara, K. Kobayashi and T. Itou, Current-induced magnetization caused by crystal chirality in nonmagnetic elemental tellurium, Phys. Rev. Res. **3**, 023111 (2021).
[29] S. Wimmer, K. Chadova, M. Seemann, D. Ködderitzsch and H. Ebert, Fully relativistic description of spin-orbit torques by means of linear response theory, Phys. Rev. B **94**, 054415 (2016).
[30] H. Grimmer, General connections for the form of property tensors in the 122 Shubnikov point groups, Acta Crystallogr. A **47**, 226 (1991).
[31] C. Wang, Y. Gao and D. Xiao, Intrinsic nonlinear Hall effect in antiferromagnetic tetragonal CuMnAs, Phys. Rev. Lett. **127**, 277201 (2021).
[32] W. H. Kleiner, Space-time symmetry of transport coefficients, Phys. Rev. **142**, 318 (1966).

[33] S. V. Gallego, J. Etxebarria, L. Elcoro, E. S. Tasci and J. M. Perez-Mato, Automatic calculation of symmetry-adapted tensors in magnetic and non-magnetic materials: a new tool of the Bilbao Crystallographic Server, Acta Crystallogr. A **75**, 438 (2019).

[34] L. Šmejkal, L. Šmejkal, L. Šmejkal, R. González-Hernández, R. González-Hernández, T. Jungwirth, T. Jungwirth, J. Sinova and J. Sinova, Crystal time-reversal symmetry breaking and spontaneous Hall effect in collinear antiferromagnets, Sci. Adv. **6**, eaaz8809 (2020).

[35] R. R. Birss, Macroscopic symmetry in space-time, Rep. Prog. Phys. **26**, 307-360 (1963).

[36] M. Yatsushiro, H. Kusunose and S. Hayami, Multipole classification in 122 magnetic point groups for unified understanding of multiferroic responses and transport phenomena, Phys. Rev. B **104**, 054412 (2021).

[37] T. Hahn, International tables for crystallography. Volume A, Space-group symmetry, (International Union of Crystallography by Kluwer Academic Publishers, Dordrecht, 1996).

[38] E. H. Hall, On a new action of the magnet on electric currents, Am. J. Math **2**, 287 (1879).

[39] R. Karplus and J. M. Luttinger, Hall effect in ferromagnetics, Phys. Rev. **95**, 1154 (1954).

[40] H. Chen, Q. Niu and A. H. MacDonald, Anomalous hall effect arising from noncollinear antiferromagnetism, Phys. Rev. Lett. **112**, 017205 (2014).

[41] W. Feng, G.-Y. Guo, J. Zhou, Y. Yao and Q. Niu, Large magneto-optical Kerr effect in noncollinear antiferromagnets $Mn_3X$ (X=Rh, Ir, Pt), Phys. Rev. B **92**, 144426 (2015).

[42] J. Liu and L. Balents, Anomalous Hall effect and topological defects in antiferromagnetic Weyl semimetals: $Mn_3Sn/Ge$, Phys. Rev. Lett. **119**, 087202 (2017).

[43] S. Nakatsuji, N. Kiyohara and T. Higo, Large anomalous Hall effect in a non-collinear antiferromagnet at room temperature, Nature **527**, 212-215 (2015).

[44] A. K. Nayak, J. E. Fischer, Y. Sun, B. Yan, J. Karel, A. C. Komarek, C. Shekhar, N. Kumar, W. Schnelle, J. Kübler, C. Felser and S. S. P. Parkin, Large anomalous Hall effect driven by a nonvanishing Berry curvature in the noncolinear antiferromagnet $Mn_3Ge$, Sci. Adv. **2**, e1501870 (2016).

[45] B. E. Zuniga-Cespedes, K. Manna, H. M. L. Noad, P.-Y. Yang, M. Nicklas,
C. Felser, A. P. Mackenzie, and C. W Hick, Observation of an anomalous Hall effect in single-crystal $Mn_3Pt$, New J. Phys. **25**, 023029 (2023).

[46] N. Kanazawa, Y. Onose, T. Arima, D. Okuyama, K. Ohoyama, S. Wakimoto, K. Kakurai, S. Ishiwata and Y. Tokura, Large topological hall effect in a short-period helimagnet MnGe, Phys. Rev. Lett. **106**, 156603 (2011).

[47] A. Neubauer, C. Pfleiderer, B. Binz, A. Rosch, R. Ritz, P. G. Niklowitz and P. Boni, Topological Hall effect in the a phase of MnSi. Phys. Rev. Lett. **102**, 186602 (2009).

[48] A. Tiwari, F. Chen, S. Zhong, E. Drueke, J. Koo, A. Kaczmarek, C. Xiao, J. Gao, X. Luo, Q. Niu, Y. Sun, B. Yan, L. Zhao and A. W. Tsen, Giant c-axis nonlinear anomalous Hall effect in $T_d$-$MoTe_2$ and $WTe_2$, Nat. Commun. **12**, 2049 (2021).

[49] Q. Ma, S.-Y. Xu, H. Shen, D. MacNeill, V. Fatemi, T.-R. Chang, A. M. Mier Valdivia, S. Wu, Z. Du, C.-H. Hsu, S. Fang, Q. D. Gibson, K. Watanabe, T. Taniguchi, R. J. Cava, E. Kaxiras, H.-Z. Lu, H. Lin, L. Fu, N. Gedik and P. Jarillo-Herrero, Observation of the nonlinear Hall effect under time-reversal-symmetric conditions, Nature **565**, 337 (2019).

[50] J. Son, K.-H. Kim, Y. H. Ahn, H.-W. Lee and J. Lee, Strain engineering of the Berry curvature dipole and valley magnetization in monolayer $MoS_2$, Phys. Rev. Lett. **123**, 036806 (2019).

[51] K. F. Mak, K. L. McGill, J. Park and P. L. McEuen, The valley hall effect in $MoS_2$ transistors, Science **344**, 1489 (2014).

[52] V. Baltz, A. Manchon, M. Tsoi, T. Moriyama, T. Ono and Y. Tserkovnyak, Antiferromagnetic spintronics, Rev. Mod. Phys. **90**, 015005 (2018).

[53] S. V. Gallego, J. M. Perez-Mato, L. Elcoro, E. S. Tasci, R. M. Hanson, K. Momma, M. I. Aroyo and G. Madariaga, MAGNDATA: towards a database of magnetic structures. I. The commensurate case, J. Appl. Crystallogr. **49**, 1750 (2016).

[54] L. Šmejkal, J. Sinova and T. Jungwirth, Emerging research landscape of altermagnetism, Phys. Rev. X **12**, 040501 (2022).

[55] L. Šmejkal, J. Sinova and T. Jungwirth, Beyond conventional ferromagnetism and antiferromagnetism: a phase with nonrelativistic spin and crystal rotation symmetry, Phys. Rev. X **12**, 031042 (2022).

[56] T. Furukawa, Y. Shimokawa, K. Kobayashi and T. Itou, Observation of current-induced bulk magnetization in elemental tellurium, Nat. Commun. **8**, 954 (2017).

[57] L. Ding, X. Xu, H. O. Jeschke, X. Bai, E. Feng, A. S. Alemayehu, J. Kim, F.-T. Huang, Q. Zhang, X. Ding, N. Harrison, V. Zapf, D. Khomskii, I. I. Mazin, S.-W. Cheong and H. Cao, Local-Ising-type magnetic order and metamagnetism in the rare-earth pyrogermanate $Er_2Ge_2O_7$, Nat. Commun. **12**, 5339 (2021).

[58] K. M. Tadde, L. Sanjeewa, J. W. Kolis, A. S. Sefat, C. delaCruz and D. M. Pajerowski, Phys. Rev. Mater. **3**, 014405 (2019).

[59] K. Kimura, Y. Kato, S. Kimura, Y. Motome and T. Kimura, Crystal-chirality-dependent control of magnetic domains in a time-reversal-broken antiferromagnet, NPJ Quantum Mater. **6**, 54 (2021).

[60] M. Ikhlas, S. Dasgupta, F. Theuss, T. Higo, S. Kittaka, B. J. Ramshaw, O. Tchernyshyov, C. W. Hicks and S. Nakatsuji, Piezomagnetic switching of the anomalous Hall effect in an antiferromagnet at room temperature, Nat. Phys. **18**, 1086 (2022).

[61] R. A. Erickson, Neutron diffraction studies of antiferromagnetism in manganous fluoride and some isomorphous compounds, Phys. Rev. **90**, 779-785 (1953).

[62] A. S. Disa, M. Fechner, T. F. Nova, B. Liu, M. Först, D. Prabhakaran, P. G. Radaelli and A. Cavalleri, Polarizing an antiferromagnet by optical engineering of the crystal field, Nat. Phys. **16**, 937-941 (2020).

[63] C.-P. Zhang, X.-J. Gao, Y.-M. Xie, H. C. Po and K. T. Law, Higher-order nonlinear anomalous Hall effects induced by Berry curvature multipoles, Phys. Rev. B **107**, 115142 (2023).

[64] W. Opechowski, Int. J. Magn. **5**, 317-325 (1974).

[65] P. Fabrykiewicz, R. Przeniosło and I. Sosnowska, Magnetic, electric and toroidal polarization modes describing the physical properties of crystals. $NdFeO_3$ case, Acta crystallogr. A **79** 80-94 (2023).

[66] H. Wang and X. Qian, Ferroelectric nonlinear anomalous Hall effect in few-layer $WTe_2$, NPJ Comput. Mater. **5**, 119 (2019).

[67] V. M. Edelstein, Spin polarization of conduction electrons induced by electric current in two-dimensional asymmetric electron systems, Solid State Commun. **73**, 233-235 (1990).

[68] H. Xu, J. Zhou, H. Wang and J. Li, Light-induced static magnetization: nonlinear Edelstein effect, Phys. Rev. B **103**, 205417 (2021).

[69] N. Jiang, Y. Nii, H. Arisawa, E. Saitoh and Y. Onose, Electric current control of spin helicity in an itinerant helimagnet, Nat. Commun. **11**, 1601 (2020).

[70] Y. Zhou and F. Liu, Realization of an antiferromagnetic superatomic graphene: Dirac Mott insulator and circular dichroism Hall effect, Nano Lett. **21**, 230-235 (2021).

[71] D. M. Juraschek, M. Fechner, A. V. Balatsky and N. A. Spaldin, Dynamical multiferroicity, Phys. Rev. Mater. **1**, 014401 (2017).
[72] E. L. Ivchenko and G. E. Pikus, New photogalvanic effect in gyrotropic crystals, JETP Lett. **27**, 604-608 (1978).
[73] V. I. Belinicher, Space-oscillating photocurrent in crystals without symmetry center, Phys. Lett. A **66**, 213-214 (1978).
[74] M. V. Hobden, Optical activity in a non-enantiomorphous crystal: $AgGaS_2$. Acta Crystallogr. A **24** (6), 676-680 (1968).
[75] P. Önnerud, Y. Andersson, R. Tellgren and P. Nordblad, The magnetic structure of ordered cubic $Pd_3Mn$, J. Solid State Chem. **128**, 109-114 (1997).
[76] G. Courbion and M. Leblanc, The magnetic structure of $NaMnFeF_6$, J. Magn. Magn. Mater. **74**, 158-164 (1988).
[77] M. Reehuis, C. Ulrich, A. Maljuk, C. Niedermayer, B. Ouladdiaf, A. Hoser, T. Hofmann and B. Keimer, Neutron diffraction study of spin and charge ordering in $SrFeO_{3-\delta}$. Phys. Rev. B **85**, 184109 (2012).
[78] S. K. Karna, D. Tristant, J. K. Hebert, G. Cao, R. Chapai, W. A. Phelan, Q. Zhang, Y. Wu, C. Dhital, Y. Li, H. B. Cao, W. Tian, C. R. DelaCruz, A. A. Aczel, O. Zaharko, A. Khasanov, M. A. McGuire, A. Roy, W. Xie, D. A. Browne, I. Vekhter, V. Meunier, W. A. Shelton, P. W. Adams, P. T. Sprunger, D. P. Young, R. Jin and J. F. DiTusa, Helical magnetic order and Fermi surface nesting in noncentrosymmetric ScFeGe, Phys. Rev. B **103**, 014443 (2021).
[79] S. Hayashida, O. Zaharko, N. Kurita, H. Tanaka, M. Hagihala, M. Soda, S. Itoh, Y. Uwatoko and T. Masuda, Pressure-induced quantum phase transition in the quantum antiferromagnet $CsFeCl_3$. Phys. Rev. B **97**, 140405 (2018).
[80] P. Lacorre, J. Pannetier, T. Fleischer, R. Hoppe and G. Ferey, Ordered magnetic frustration: XVI. Magnetic structure of $CsCoF_4$ at 1.5 K. J. Solid State Chem. **93**, 37-45 (1991).
[81] K. Ishito, H. Mao, Y. Kousaka, Y. Togawa, S. Iwasaki, T. Zhang, S. Murakami, J.-I. Kishine and T. Satoh, Truly chiral phonons in α-HgS, Nat. Phys. **19**, 35-39 (2023)
[82] H. Ueda, M. García-Fernández, S. Agrestini, C. P. Romao, J. van den Brink, N. A. Spaldin, K.-J.Zhou and U. Staub, Chiral phonons in quartz probed by X-rays, Nature **618**, 946-950 (2023)
[83] S. N. Achary and A. K. Tyagi, Strong anisotropic thermal expansion in cristobalite-type $BPO_4$, J. Solid State Chem. 177, 3918-3926 (2004)
[84] J. B. Parise and T. E. Gier, Hydrothermal syntheses and structural refinements of single crystal lithium boron germanate and silicate, $LiBGeO_4$ and $LiBSiO_4$, Chem. Mater. **4**, 1065-1067 (1992)
[85] R. Diehl and C.-D. Carpentier, The structural chemistry of indium phosphorus chalcogenides, Acta. Cryst. B **34**, 1097-1105 (1978)
[86] Bindi, L., Bonazzi, P., Dusek, M., Petricek, V. and Chapuis, Five-dimensional structure refinement of natural melilite, $(Ca_{1.89}Sr_{0.01}Na_{0.08}K_{0.02})(Mg_{0.92}Al_{0.08})$-$(Si_{1.98}Al_{0.02})O_7$, Acta. Cryst. B **57**, 739-746 (2001)
[87] Yamaoka, S., Lemley, J.T., Jenks, J.M., Steinfink, H. Structural chemistry of the polysulfides dibarium trisulfide and monobarium trisulfide, Inorg. Chem. **14**, 129-131 (1975)